\documentclass[twocolumn,english,pra]{revtex4-2}
\usepackage[T1]{fontenc}
\usepackage[latin9]{inputenc}
\usepackage{color}
\usepackage{babel}
\usepackage{amsmath}
\usepackage{amssymb}
\usepackage{esint}
\usepackage{graphicx}
\usepackage{hyperref}
\hypersetup{
           breaklinks=true,   
           colorlinks=true,   
           pdfusetitle=true,  
}

\makeatletter

\usepackage{braket}
\usepackage{babel}

\begin{document}
\title{Dynamics of the molecular geometric phase}
\author{Rocco Martinazzo$^{1,2,*}$, Irene Burghardt$^{3}$}
\affiliation{$^{1}$Department of Chemistry, Università degli Studi di Milano,
Via Golgi 19, 20133 Milano, Italy}
\email{rocco.martinazzo@unimi.it}

\affiliation{$^{2}$Istituto di Scienze e Tecnologie Molecolari, CNR, via Golgi
19, 20133 Milano, Italy}
\affiliation{$^{3}$Institute of Physical and Theoretical Chemistry, Goethe University
Frankfurt, Max-von-Laue-Str. 7, D-60438 Frankfurt/Main, Germany}
\begin{abstract}
The fate of the molecular geometric phase in an exact dynamical framework
is investigated with the help of the exact factorization of the wavefunction
and a recently proposed quantum hydrodynamical description of its
dynamics. An instantaneous, gauge invariant phase is introduced for
arbitrary paths in nuclear configuration space in terms of hydrodynamical
variables, and shown to reduce to the adiabatic geometric phase when
the state is adiabatic and the path is closed. The evolution of the
closed-path phase over time is shown to adhere to a Maxwell-Faraday
induction law, with non-conservative forces arising from the electron
dynamics that play the role of electromotive forces. We identify the
pivotal forces that are able to change the value of the phase, thereby
challenging any topological argument. Nonetheless, negligible changes
in the phase occur when the local dynamics along the probe loop is
approximately adiabatic. In other words, the adiabatic idealization
of geometric phase effects may remain suitable for effectively describing
certain dynamic observables. 
\end{abstract}
\maketitle
\textbf{\emph{Introduction}}. Geometric phases are fundamental concepts
in physics and chemistry, with wide-ranging implications. They are
closely associated with various phenomena, such as the quantum, the
anomalous and the spin Hall effect \citep{Girvin2019,VanderbiltBook2018},
the exotic physics of topological insulators \citep{Hasan2010,Kane2013},
dielectric polarization in crystals \citep{Resta1992b,King-Smith1993b,Resta1994b,Resta2000,Xiao2010,VanderbiltBook2018},
the Aharonov-Bohm effect \citep{Aharonov1959}, and conical intersections
(CIs) in molecules \citep{Mead1992,Yarkony1996,Kendrick2003}. Geometric
phases usually emerge when the Hamiltonian of a system depends on
a set of ``environmental'' parameters $\mathbf{x}$ which, in Berry's
original work \citep{Berry1984}, are allowed to change adiabatically,
 but they remain well defined concepts for non-adiabatic, non-cyclic
and non-unitary evolutions as well \citep{Simon1983,Aharonov1987,Simon1993,Mukunda1993}.
 In the case of molecules, geometric phases play a critical role
around an intersection between two or more potential energy surfaces.
Even when the molecular dynamics remains nearly adiabatic, the presence
of a CI can significantly impact the outcome of a chemical reaction
\citep{Juanes-Marcos2005,Yuan2020,Kendrick2015}, because of the
quantum interference of wavepackets encircling the CI that crucially
depends on the geometric phase \citep{Althorpe2006,Valahu2023}. In
these molecular problems, the Berry phase is often not just geometric
but also topological, that is, it is independent of both the dynamics
and the path (as long as homotopic paths are compared). In fact,
it is the phase introduced as early as 1958 by Longuet-Higgins \citep{Longuet-Higgins1958,Herzberg1963}
that is known to control the energy level ordering in, e.g., Jahn-Teller
systems. However, these intriguing properties depend crucially on
the adiabatic description (approximation) of the dynamics, and it
is uncertain whether and how they persist when the \emph{exact} quantum
dynamics is considered. Recent works \citep{Min2014,Requist2016a}
have shown that the topological character of the phase is an artifact
of the adiabatic approximation and suggest, more generally, that the
geometric phase in molecules may be a less useful concept than previously
believed. 

The purpose of this work is to shed light on these issues and to reconcile
the adiabatic perspective with the exact dynamical evolution. To this
end we will first show that a geometric phase is yet meaningful when
the full electron-nuclear system is in a pure state. We will use the
framework of the exact factorization (EF) of the molecular wavefunction
\citep{Abedi2010,Abedi2012} since this construction extends the fiber
structure of the adiabatic approximation to arbitrary states, thereby
enabling a natural extension of the Berry phase \footnote{Henceforth, we shall use the term Berry phase to mean the ``adiabatic
geometric phase''}. Subsequently, we will explore the exact dynamical evolution of this
phase. This task is challenging when using the original equations
of motion of the EF approach due to their inherent \emph{gauge} freedom.
However, a recently developed quantum hydrodynamical (QHD) description
of the EF dynamics \citep{Martinazzo2023} makes this step feasible.
QHD offers an alternative formulation for electron-nuclear dynamics,
relying on EF while employing only \emph{gauge}-invariant variables
\citep{Martinazzo2023}.  Within this QHD-EF framework we will identify
the key factors influencing the evolution of the geometric phase and
analyze a model two-state problem. 

\textbf{\emph{Gauge}}\textbf{ }\textbf{\emph{invariant EF dynamics}}.
In the exact factorization approach \citep{Abedi2010,Abedi2012} the
wavefunction is represented exactly as 
\begin{equation}
\ket{\Psi}=\int_{X}d\mathbf{x}\psi(\mathbf{x})\ket{u(\mathbf{x})}\ket{\mathbf{x}}\label{eq:EF wavefunction}
\end{equation}
where $\{\ket{\mathbf{x}}\}$ is the position basis of the nuclear
variables $x^{k}$ ($k=1,2,..N$), $\ket{u(\mathbf{x})}$ is the conditional
electronic state at $\mathbf{x}$ and $\psi(\mathbf{x})$ is the marginal
probability amplitude for the nuclei, i.e. the ``nuclear wavefunction''.
The latter two can be obtained, up to a \emph{gauge} choice, by projecting
the total wavefunction on the nuclear basis states and imposing a
normalization condition on the ensuing local electronic states,
\begin{equation}
\braket{\mathbf{x}|\Psi}_{X}=\psi(\mathbf{x})\ket{u(\mathbf{x})}\equiv\ket{\boldsymbol{\Psi}(\mathbf{x})}\ \braket{u(\mathbf{x})|u(\mathbf{x})}=1\label{eq:EF definition}
\end{equation}
where the subscript $X$ indicates that integration is performed over
nuclear variables only. In the EF approach $\psi(\mathbf{x})$ and
$\ket{u(\mathbf{x})}$ evolve in time according to equations of motion
that can be derived from either the variational principle or projection
operator techniques, as shown, respectively, in Refs. \citep{Abedi2010,Abedi2012}
or Refs. \citep{Martinazzo2022,Martinazzo2022a}. In the QHD description
of the dynamics \citep{Martinazzo2023} the nuclear wavefunction is
replaced by a probability fluid with density $n(\mathbf{x})=|\psi(\mathbf{x})|^{2}$
and velocity field $v^{k}(\mathbf{x})$, and the electronic state
is described by electronic density operators $\rho_{\text{el}}(\mathbf{x})=\ket{u(\mathbf{x})}\bra{u(\mathbf{x})}$
tied to the fluid elements. The equations of motion consist of a continuity
equation for the density, $\partial_{t}n+\sum_{k}\partial_{k}(nv^{k})=0$,
a momentum equation 
\begin{align}
\dot{\pi}_{k} & =-\text{Tr}_{\text{el}}(\rho_{\text{el}}\partial_{k}H_{\text{el}})-\frac{\hbar^{2}}{n}\sum_{ij}\xi^{ij}\partial_{i}(ng_{kj})-\partial_{k}Q\label{eq:momentum equation}
\end{align}
and a Liouville-von Neumann-like equation for the conditional electronic
density operators,
\begin{equation}
i\hbar\dot{\rho}_{\text{el}}=[H_{\text{el}}+\delta H_{\text{en}},\rho_{\text{el}}]\label{eq:LvN equation for electrons}
\end{equation}
In the above equations the dot denotes the material (i.e., Lagrangian)
derivative, $\dot{f}\equiv\partial_{t}f+\sum_{k}v^{k}\partial_{k}f$,
$\xi^{ij}$ is the inverse mass tensor of the nuclear system, $Q=-\hbar^{2}/2\sum_{ij}\xi^{ij}n^{-1/2}\partial_{i}\partial_{j}n^{1/2}$
is the Bohm quantum potential \citep{Holland1993}, $H_{\text{el}}=H_{\text{el}}(\mathbf{x})$
is the local electronic Hamiltonian and $\delta H_{\text{en}}$ is
the electron-nuclear coupling 
\begin{equation}
\delta H_{\text{en}}=-\frac{\hbar^{2}}{2n}\sum_{ij}\xi^{ij}\partial_{i}(n\partial_{j}\rho_{\text{el}})\label{eq:e-n coupling Hamiltonian}
\end{equation}
Furthermore, $g_{kj}$ is the Fubini-Study (FS) metric \citep{Provost1980},
which is the real part of the quantum geometric tensor \citep{Berry1984}
\begin{equation}
q_{kj}=\text{Tr}_{\text{el}}(\rho_{\text{el}}\partial_{k}\rho_{\text{el}}\partial_{j}\rho_{\text{el}})\label{eq:quantum geometric tensor}
\end{equation}
here expressed in terms of conditional density operators. The momentum
field $\pi_{k}$ is related to the velocity field through the inverse
mass tensor, $v^{k}=\sum_{j}\xi^{kj}\pi_{j}$, and is connected to
the EF wavefunction by $\pi_{k}=\Re(\hat{p}_{k}\psi/\psi)-\hbar A_{k}$,
where $\hat{p}_{k}=-i\hbar\partial_{k}$ is the canonical momentum
operator and $A_{k}$ is the Berry connection, $A_{k}=i\braket{u|\partial_{k}u}$.
It can also be obtained from the total electronic-nuclear ($e-n$)
wavefunction, without refererring to the EF, since $\pi_{k}\equiv n^{-1}(\mathbf{x})\Re\braket{\boldsymbol{\Psi}(\mathbf{x})|\hat{p}_{k}|\boldsymbol{\Psi}(\mathbf{x})}_{\text{el}}$,
where $\ket{\boldsymbol{\Psi}(\mathbf{x})}$ was introduced in Eq.
\ref{eq:EF definition} and the subscript $\text{el}$ means that
integration is performed over electronic variables only. The circulation
of $\pi_{k}$ around arbitrary closed paths $\gamma$ in nuclear configuration
space satisfies a quantization condition
\begin{equation}
\sum_{k}\oint_{\gamma}(\pi_{k}+\hbar A_{k})dx^{k}=2\pi\hbar\,n\ \ n\in\mathbb{Z}\label{eq:quantization condition}
\end{equation}
which merely expresses the fact that the nuclear wavefunction $\psi$
must be smooth around any loop. Here, $n$ is the topological value
that describes the way the wavefunction phase winds around a singularity
of the momentum field (for instance, a wavefunction node). In fact,
$n$ can be nonzero only in a multiply connected domain, when $\gamma$
cannot be shrunk to a single point. The condition of Eq. \ref{eq:quantization condition}
needs to imposed at the initial time only (with a smooth choice of
the phases of the nuclear and electronic wavefunctions in Eq. \ref{eq:EF definition}),
since Kelvin's circulation theorem holds for the fluid dynamics described
here \citep{Martinazzo2023}. Hence, in principle, the EF nuclear
and electronic wavefunctions are needed only at the initial time,
and $n$, $\pi_{k}$ and $\rho_{\text{el}}$ can then be obtained
at any time upon solving the above \emph{gauge}-invariant equations
of motion, i.e., Eqs. \ref{eq:momentum equation}, \ref{eq:LvN equation for electrons}
jointly with the continuity equation. In the following we shall focus
on a given instant of time and investigate the geometric properties
of the instantaneous fiber bundle induced by the EF of the total wavefunction. 

\textbf{\emph{Non-adiabatic geometric phase}}. The quantization condition
of Eq. \ref{eq:quantization condition} shows that the momentum field
$\pi_{k}$ can be used to define a \emph{gauge} invariant, instantaneous
``phase'' for arbitrary paths $\gamma$ in configuration space
\begin{equation}
\Gamma[\gamma]=-\frac{1}{\hbar}\sum_{k}\int_{\gamma}\pi_{k}dx^{k}\label{eq:definition of QHD phase}
\end{equation}
For closed paths this reduces, by construction, to the holonomy of
the vector bundle defined by the EF representation of the wavefunction
(see Eq. \ref{eq:quantization condition})
\begin{equation}
\Gamma_{\text{el}}[\gamma]=\sum_{k}\oint_{\gamma}A_{k}dx^{k}\ \ \ (\text{mod}2\pi)\label{eq:Berry phase for loops}
\end{equation}
and, more generally, it defines a quantity which is \emph{gauge} invariant
even when the path is open. In fact, for a curve $\gamma$ that connects
$\mathbf{x}_{a}$ to $\mathbf{x}_{b}$, $\Gamma[\gamma]$ is the sum
of two contributions that are separately invariant, 
\begin{equation}
\Gamma[\gamma]=-\Theta_{ba}+\Gamma_{\text{el}}[\gamma]\label{eq:phase decomposition}
\end{equation}
where $\Theta_{ba}=\text{arg}\braket{\boldsymbol{\Psi}(\mathbf{x}_{a})|\boldsymbol{\Psi}(\mathbf{x}_{b})}$
and $\Gamma_{\text{el}}[\gamma]$ is a purely electronic term 
\begin{equation}
\Gamma_{\text{el}}[\gamma]=\text{arg}\braket{u(\mathbf{x}_{a})|u(\mathbf{x}_{b})}+\sum_{k}\int_{\gamma}A_{k}dx^{k}\label{eq:open-path Berry phase}
\end{equation}
The first term, $\Theta_{ba}$, is the Pancharatnam phase difference
(valid for arbitrary non-orthogonal vectors) of the total electron-nuclear
wavefunction between $b$ and $a$, while the second term, $\Gamma_{\text{el}}[\gamma]$
is the Pancharatnam phase accumulated by the electronic vector when
parallel transported from $\mathbf{x}_{a}$ to $\mathbf{x}_{b}$ along
$\gamma$ \citep{Simon1983,Simon1993,Mukunda1993,Pati1995,Pati1998}\footnote{The underlying connection being, of course, the instantaneous Berry
connection defined by the EF of the wavefunction.}. Indeed, $\pi_{k}\equiv\hbar\partial_{k}\theta-\hbar A_{k}$ (where
$\theta=\text{arg}\psi$) and the phase change of the nuclear wavefunction
at the endpoints of the curve, $\Delta\theta=\theta_{b}-\theta_{a}$,
can be written as $\Delta\theta=\text{arg}(\psi_{a}^{*}\psi_{b}\braket{u_{a}|u_{b}})-\text{arg}(\braket{u_{a}|u_{b}})$,
where $\psi(\mathbf{x})\ket{u(\mathbf{x})}=\ket{\boldsymbol{\Psi}(\mathbf{x})}$.
In a sense, $-\Gamma[\gamma]$ is a \emph{nuclear} phase, i.e., the
phase difference of the total wavefunction minus that of the electronic
one. That is, $\Theta_{ba}=-\Gamma[\gamma]+\Gamma_{\text{el}}[\gamma]$
is a decomposition of the total phase difference into nuclear and
electronic contributions. For a loop $\Theta_{ba}\equiv0$ and the
nuclear phase difference is the opposite of the electronic one, which
in this case reduces to Eq. \ref{eq:Berry phase for loops}.

Importantly, $\Gamma[\gamma]$ is only indirectly tied to the connection
defined by the EF: it is a property that relies on EF but does not
require that the EF of the wavefunction is performed. Its definition
is further consistent with the fluid dynamics: for stationary loops
the quantization condition (Eq. \ref{eq:quantization condition})
jumps eventually by $\pm2\pi$ during the dynamics (every time a wavefunction
node crosses the loop \citep{Martinazzo2023}) but this does not affect
the above identities.

\textbf{\emph{Dynamics}}. We now focus on the phase defined by Eq.
\ref{eq:definition of QHD phase}, evaluated for a path $\gamma$
that is a loop, fixed in time, and use the symbol $\Gamma_{O}[\gamma]$
to emphasize that the path is closed. The dynamical evolution of $\Gamma_{O}[\gamma]$
is determined by the momentum equation, upon observing that $\partial_{t}\pi_{k}=\dot{\pi}_{k}-\sum_{j}v^{j}\partial_{j}\pi_{k}$.
Here, the first term is the force acting on the fluid elements (Eq.
\ref{eq:momentum equation}) and the second one (the advective contribution)
can be rearranged as $\sum_{j}v^{j}\partial_{j}\pi_{k}=\sum_{j}v^{j}\partial_{k}\pi_{j}+\sum_{j}v^{j}B_{kj}$,
where only the second term contributes to $\Gamma_{O}[\gamma]$, since
the first one is the $k^{\text{th}}$ derivative of the classical
kinetic energy $T=\sum_{ij}\xi^{ij}\pi_{i}\pi_{j}$. Furthermore,
$B_{kj}$ is the $kj$ component of the curvature tensor $B_{kj}=\hbar(\partial_{k}A_{j}-\partial_{j}A_{k})=-2\hbar\Im q_{kj}$,
with the result that the rate of change of the phase displays three
distinct, \emph{gauge}-invariant contributions,
\begin{equation}
-\frac{d\Gamma_{O}[\gamma]}{dt}=\mathfrak{E}^{\text{NBO}}+\mathfrak{E}^{\text{el}}+\mathfrak{E^{\text{mag}}}\label{eq:rate of phase change - fixed loop}
\end{equation}
Here, 
\begin{equation}
\mathfrak{E}^{\text{NBO}}=-\frac{1}{\hbar}\oint_{\gamma}\sum_{k}\text{Tr}_{\text{el}}(\rho_{\text{el}}\partial_{k}H_{\text{el}})dx^{k}\label{eq:NBO circulation}
\end{equation}
is a non-adiabatic contribution driven by the electronic Hamiltonian,
and
\begin{equation}
\mathfrak{E}^{\text{el}}=-\hbar\oint_{\gamma}\sum_{ijk}\frac{\xi^{ij}}{n}\partial_{i}(ng_{kj})dx^{k}\label{eq:pseudo-electric cicurlation}
\end{equation}
\begin{equation}
\mathfrak{E}^{\text{mag}}=-\frac{1}{\hbar}\oint_{\gamma}\sum_{jk}v^{j}B_{kj}dx^{k}\label{eq:pseudo-magentic circulation}
\end{equation}
are geometric contributions related, respectively, to the \emph{pseudo}-electric
and \emph{pseudo}-magnetic \emph{gauge} fields acting on the nuclei. 

Eq. \ref{eq:NBO circulation} represents a genuine non-Born-Oppenheimer
contribution entirely due to the non-conservative part of the Ehrenfest
force $F_{k}^{\text{Eh}}=-\text{Tr}_{\text{el}}(\rho_{\text{el}}\partial_{k}H_{\text{el}})$
appearing in Eq. \ref{eq:momentum equation}. It is tied to the non-stationarity
of the local electronic states since it disappears when the system
is in an adiabatic state, i.e. when setting $\ket{u(\mathbf{x})}$
to be eigenstate of the local electronic Hamiltonian $H_{\text{el}}(\mathbf{x})$.
The second contribution, Eq. \ref{eq:pseudo-electric cicurlation},
is generally non-vanishing, and it depends on the (instantaneous)
electronic state through the metric properties of the EF fiber bundle
(the FS metric $g_{kj}$) and on the nuclear state through the density
$n$. The third contribution, Eq. \ref{eq:pseudo-magentic circulation},
on the other hand, is (possibly) non-vanishing only when the nuclear
state is current-carrying. It appears here only because we fixed the
loop: the phase is tied to the local electronic states that, in turn,
move in tandem with the fluid elements describing the nuclear probability
density. In other words, if we allowed the loop to follow the fluid
dynamics we would find
\begin{equation}
-\frac{d\tilde{\Gamma}_{O}[\gamma]}{dt}=\mathfrak{E}^{\text{NBO}}+\mathfrak{E}^{\text{el}}\label{eq:rate of phase change - dynamical loop}
\end{equation}
where now $\tilde{\Gamma}_{O}[\gamma]$ refers to the geometric phase
along a loop $\gamma$ that follows the fluid flow (see SM). 

The above findings are general, and hold for arbitrary electronic-nuclear
states. For cases where Stokes's theorem applies they can be anticipated
by the Maxwell-Faraday induction law
\begin{equation}
-\partial_{t}\mathcal{B}=d\mathcal{E}\label{eq:Maxwell-Faraday law}
\end{equation}
that holds for the \emph{gauge} fields governing the nuclear dynamics
in the EF approach \citep{Martinazzo2023}. Here, $d$ denotes the
exterior derivative, $\mathcal{B}=d\omega$ is the Berry curvature
2-form, $\omega=\hbar\sum_{k}A_{k}dx^{k}$ is the 1-form associated
with the Berry connection, and $\mathcal{E}=i\hbar d\braket{u|\partial_{t}u}-\partial_{t}\omega=\sum_{k}E_{k}dx^{k}$
is the \emph{gauge}-invariant 1-form defining the \emph{pseudo-}electric
field $E_{k}$ \footnote{There is some freedom in defining this field, since one can add any
\emph{gauge}-invariant closed form to $\mathcal{E}$ without altering
the induction law and its \emph{gauge} invariance. For instance, one
can choose to add conservative forces in the form $\mathcal{F}=d\mathcal{V}$,
where $\mathcal{V}$ is a (\emph{gauge} invariant) scalar potential
acting on the nuclei. The important, non conservative, part of $\mathcal{E}$
is the electron dynamical force $F_{k}^{\text{ED}}$ first introduced
in Ref. \citep{Martinazzo2022}: in the adiabatic approximation $F_{k}^{\text{ED}}\equiv0$
and the geometric phase is stationary. }. Indeed, application of Stokes' theorem to an open surface having
$\gamma$ as a boundary, and identification of the \emph{pseudo}-electric
field $E_{k}$ acting on the nuclei (see Section III.A of Ref. \citep{Martinazzo2023}
and, in particular, Eq. 59) leads again to Eqs. \ref{eq:rate of phase change - fixed loop}-\ref{eq:pseudo-magentic circulation}.
Compared to the Maxwell-Faraday induction law of classical electromagnetism,
though, here there is no varying magnetic flux inducing an electromotive
force on a circuit. Rather, it is the magnetic flux (i.e. the geometric
phase) of the electronic subsystem that changes \emph{because} of
the non-conservative work done by the electrons on the nuclei, around
the loop $\gamma$ in nuclear configuration space. That is, Eq. \ref{eq:Maxwell-Faraday law}
becomes a reversed-induction law.

Of main interest here is the analysis of the adiabatic geometric phase,
i.e. the rate of phase change defined by Eq. \ref{eq:rate of phase change - fixed loop}
when the system is found in an adiabatic state and, in particular,
when the phase is topological. In this situation, as mentioned above,
the non-Born-Oppenheimer circulation of Eq. \ref{eq:NBO circulation}
vanishes since the Ehrenfest force becomes conservative. However,
also the ``drift'', \emph{pseudo}-magnetic term of Eq. \ref{eq:pseudo-magentic circulation}
disappears since $\mathcal{B}\equiv0$ (almost everywhere) if the
phase is topological. Hence, we are left with the \emph{pseudo}-electric
work of Eq. \ref{eq:pseudo-electric cicurlation} which thus represents
the key factor affecting a phase which is found topological in the
adiabatic approximation, at least at short time when departing from
the adiabatic state. Since Eq. \ref{eq:pseudo-electric cicurlation}
is generally non-zero and path-dependent, the phase which is topological
in the adiabatic state becomes geometric. Indeed, the curvature departs
from zero according to the induction law, Eq. \ref{eq:Maxwell-Faraday law},
driven by the local vorticity of the \emph{pseudo}-electric field
which is generally non-zero. Hence, $\mathcal{B}$ becomes non-zero
and the phase cannot remain topological. In the following we address
a model two-state problem that exemplifies the transition between
topological and geometric phase. 

\textbf{\emph{Model two-state problem.}} We now consider a 2-state
model that highlights the key features of a molecular problem involving
a CI. The nuclear system contains a number of degrees of freedom described
by $\mathbf{x}\in\mathcal{M}\cong\mathbb{R}^{N}$ and the electronic
Hamiltonian takes the general form (in a diabatic basis \citep{Kouppel1984})
$H_{\text{el}}=A(\mathbf{x})\text{\ensuremath{\sigma_{0}}}+\mathbf{B}(\mathbf{x})\boldsymbol{{\sigma}}$,
where $A(\mathbf{x})$ is a scalar, $\mathbf{B}(\mathbf{x})\in\mathcal{N}\cong\mathbb{R}^{3}$
is an effective magnetic field, $\sigma_{0}=\mathbb{I}_{2}$ is the
2x2 unit matrix and $\boldsymbol{{\sigma}}=(\sigma_{x},\sigma_{y},\sigma_{z})$
is the vector of Pauli matrices. 

The geometric properties of the adiabatic bundles are well-known \citep{Berry1984,Bohm2003}
and can be ``pulled back'' from the problem of a spin in a slowly
varying magnetic field, with $\mathcal{N}$ as parameter space \footnote{More precisely, the quantum geometric tensor $q$ in parameter space
$\mathcal{M}$ is the pullback by the magnetic field $\beta:\mathcal{M}\rightarrow\mathcal{N}$
of the fundamental geometric tensor $\tilde{q}$ in $\mathcal{N}$,
i.e., $q_{\mathbf{x}}(\mathbf{u},\mathbf{v})=(\beta^{*}\tilde{q})_{\mathbf{x}}(\mathbf{u},\mathbf{v})\equiv\tilde{q}_{\beta(\mathbf{x})}(d\beta_{\mathbf{x}}(\mathbf{u}),d\beta_{\mathbf{x}}(\mathbf{v}))$
for $\mathbf{x}\in\mathcal{M}$ and $\mathbf{u},\mathbf{v}\in T_{\mathbf{x}}\mathcal{M}$.
Here $\beta:\mathbf{x}\rightarrow\mathbf{B}(\mathbf{x})$ specifies
the magnetic field and $q$, $\tilde{q}$ are sections of $T^{*}\mathcal{M}\otimes T^{*}\mathcal{M}$
and $T^{*}\mathcal{N}\otimes T^{*}\mathcal{N}$, respectively.}. Briefly, the Berry phase along a path $\gamma\subset\mathcal{M}$
is generally path-dependent and is given by $q_{\pm}=\mp\hbar/2$
times the solid angle subtended by the curve image $\tilde{\gamma}=\beta\circ\gamma$
generated in $\mathcal{N}$ space by the magnetic field function,
$\beta:\mathbf{x}\rightarrow\mathbf{B}(\mathbf{x})$. This is the
flux of the ``Berry field'' $\mathbf{H}^{\pm}=q_{\pm}\mathbf{B}/B^{3}$
through an arbitrary surface subtending the curve $\tilde{\gamma}$
\footnote{In $\hbar$ units, with our convention, since we have included $\hbar$
in the definition of $\omega$.}. In a typical molecular problem, however, one of the $\mathbf{B}$
components identically vanishes because of time-reversal symmetry,
say $B_{z}$. Hence, the curvature in nuclear configuration space
becomes $\mathcal{B}_{\pm}=H_{z}^{\pm}dB^{x}\wedge dB^{y}$ and it
vanishes everywhere except at the CI seam. The image paths $\tilde{\gamma}$
necessarily lie in the $xy$ plane of the $\mathcal{N}$ space and
the Berry phase becomes $\mp n\pi$, where $n$ is the winding number
of $\tilde{\gamma}$ around the origin of $\mathcal{N}$. 

In the adiabatic approximation the magnetic field $\mathbf{B}$ fully
characterizes the electron dynamics (i.e., the energetics of the electronic
problem) and the structure of the relevant bundle. For arbitrary electron-nuclear
states (and an exact dynamics) we further need the polarization vector
$\mathbf{s}(\mathbf{x})\in\mathcal{N}$ that characterizes the conditional
density matrix, $\rho_{\text{el}}(\mathbf{x})=\frac{1}{2}(\sigma_{0}+\mathbf{s}(\mathbf{x})\boldsymbol{\sigma})$
($\lVert\mathbf{s}\rVert=1$ for a pure state). Eq. \ref{eq:LvN equation for electrons}
gives its dynamical equation in the form 
\begin{equation}
\dot{\mathbf{s}}=(\Omega\mathbf{b}+\boldsymbol{\tau})\times\mathbf{s}-\frac{\hbar}{2}\sum_{ij}\xi^{ij}\partial_{i}(\mathbf{s}_{j}\times\mathbf{s})\label{eq:2-state electron dynamics}
\end{equation}
where $\Omega=2B/\hbar$ is the Larmor precession frequency, $\mathbf{b}$
is the unit vector identifying the magnetic field direction, $\mathbf{s}_{j}:=\partial_{j}\mathbf{s}$
and $\boldsymbol{\tau}=\sum_{j}u^{j}\mathbf{s}_{j}$ is the ``nuclear
torque'', an effective field due to the \emph{e-n} coupling that
involves the osmotic velocity \footnote{Here, $w_{k}=-\hbar/2\partial_{k}\ln n$ is the imaginary component
of the complex-valued momentum field introduced in Ref. \citep{Martinazzo2023},
see Section II.A. } $u^{j}=-\hbar/2\sum_{k}\xi^{jk}\partial_{k}\ln n$. The bundle structure,
on the other hand, is characterized by the quantum geometric tensor
\begin{equation}
q_{kj}=\frac{1}{4}\left(\mathbf{s}_{k}\mathbf{s}_{j}+i\mathbf{s}(\mathbf{s}_{k}\times\mathbf{s}_{j})\right)\label{eq:2-state quantum geometric tensor}
\end{equation}
which gives $g_{kj}=\mathbf{s}_{k}\mathbf{s}_{j}/4$ and $B_{kj}=-\hbar\mathbf{s}(\mathbf{s}_{k}\times\mathbf{s}_{j})/2$
(see SM). We can therefore express the contributions to the rate of
phase change appearing on the r.h.s. of Eq. \ref{eq:rate of phase change - fixed loop}
as integrals of simple 1-forms, i.e. $\mathfrak{E}^{\text{X}}=\oint_{\gamma}\Phi^{\text{X}}$
(for X = NBO, el and mag), where $\Phi^{\text{NBO}}=\hbar^{-1}\mathbf{B}d\mathbf{s}$,
$\Phi^{\text{el}}=1/2\boldsymbol{\tau}d\mathbf{s}-\hbar/4\sum_{ij}\xi^{ij}\partial_{i}\mathbf{s}_{j}\,d\mathbf{s}$
and $\Phi^{\text{mag}}=1/2\ (\boldsymbol{\nu}\times\mathbf{s})d\mathbf{s}$,
upon defining $\boldsymbol{\nu}=\sum_{j}v^{j}\mathbf{s}_{j}$. The
circulations are therefore all mapped on the Bloch sphere $S^{2}\subset\mathcal{N}$
(the projective Hilbert space of the 2-level system), where \textbf{$\mathbf{s}(\mathbf{x})$}
traces a curve $\tilde{\gamma}$ when $\mathbf{x}$ moves along the
curve $\gamma$. Note that $\boldsymbol{\tau},\boldsymbol{\nu}$ and
$d\mathbf{s}$ are tangent to the sphere, while $\mathbf{B}$ can
have both tangent and normal components, depending on the real-space
position $\mathbf{x}$. The results for an adiabatic state follow
upon setting $\mathbf{s}=\pm\mathbf{b}$ for the upper and lower adiabatic
states, respectively. In this case, as anticipated above, $\Phi^{\text{NBO}}=0$
since $\mathbf{b}$ is normal to $S^{2}$. Furthermore, if the Berry
phase is topological, $\boldsymbol{\nu}$ and $d\mathbf{s}$ are always
parallel to each other, and we have $\Phi^{\text{mag}}=0$, as seen
above. 
\begin{figure*}
\includegraphics[width=0.42\paperwidth]{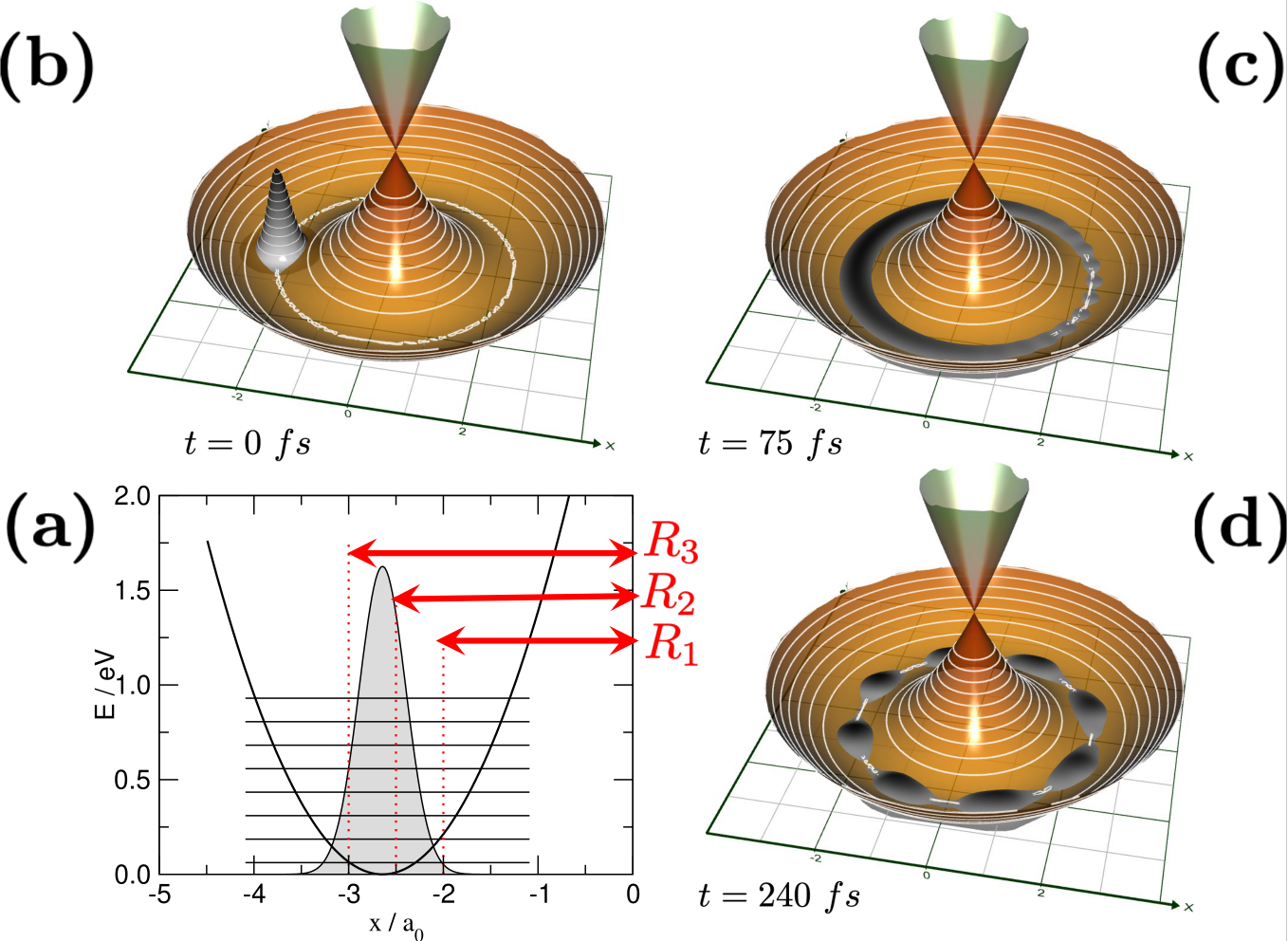}~~~~~\includegraphics[clip,width=0.38\paperwidth]{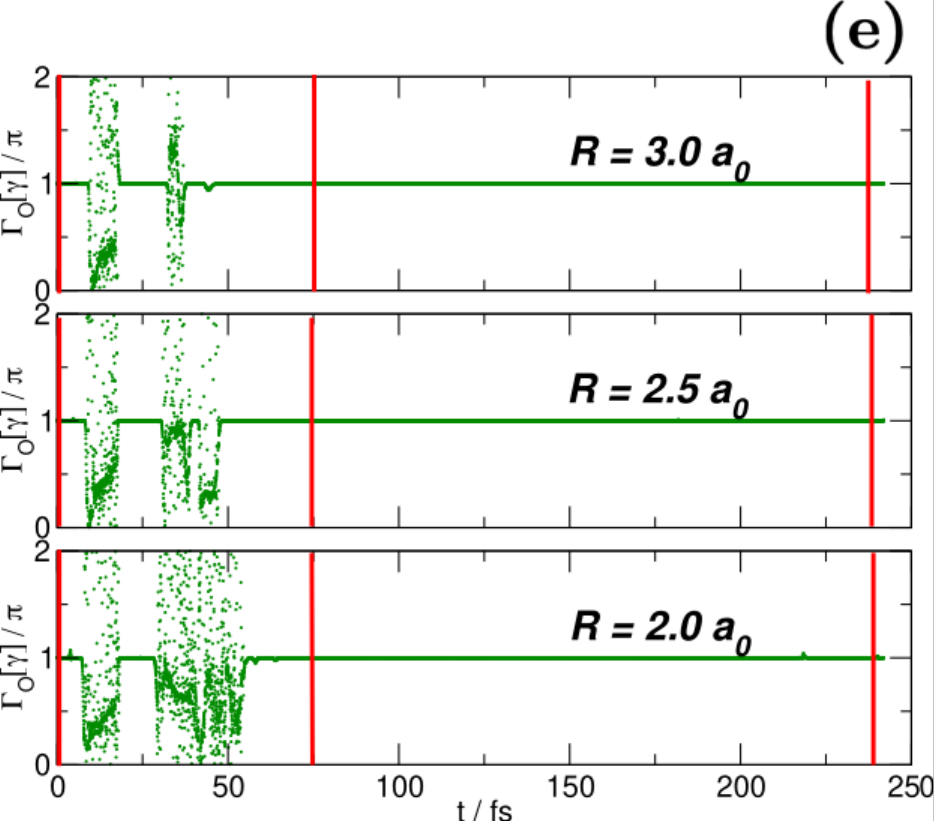}

\caption{\label{fig:Exact-quantum-dynamical results}Exact quantum dynamical
results for the two-dimensional, two-state model problem described
in the main text, as obtained from a true adiabatic ground state at
initial time. (a) Nuclear density at $t=0$ along the $x$ coordinate.
Also shown the adiabatic ground-state potential and the ladder of
vibrational states of the diabatic potential. The arrows denote the
radii of some circular paths, centered at the CI point, along which
the geometric phase was computed. (b)-(d) Snapshots of the nuclear
density at three significant times, $t=0,75$ and $240\,\,fs$, as
indicated. Also shown the adiabatic potential energy surfaces intersecting
at the origin of the coordinate system. (e) Evolution of the geometric
phase of Eq. \ref{eq:definition of QHD phase} (in units of $\pi$)
along the three paths marked in (a), namely for circles of radius
$R=2.0,2.5$ and $3.0\,a_{0}$ (from bottom to top, as indicated).
The red vertical bars denote the three times chosen for panels (b)-(d).
See text for details. }

\end{figure*}

For concreteness, we consider 2+1 nuclear degrees of freedom ($\mathcal{M}\cong\mathbb{R}^{3})$
mimicking a one-dimensional CI seam, with parameter values typical
of a molecular problem and a diagonal mass tensor, $\xi^{ij}=\delta^{ij}M^{-1}$,
with $M=1$ u.m.a.. Upon taking $A(\mathbf{x})$ independent of $z$
the problem becomes effectively two-dimensional. Specifically, setting
$A(\mathbf{x})=\frac{1}{2}M\omega_{x}^{2}x^{2}+\frac{1}{2}M\omega_{y}^{2}y^{2}$
($\omega_{x}=\omega_{y}=\omega=1000$ cm$^{-1}$) and employing a
linear vibronic coupling $\mathbf{B}=\kappa_{x}x\mathbf{e}_{1}+\text{\ensuremath{\kappa}}_{y}y\mathbf{e}_{2}$
($\kappa_{x}=\kappa_{y}=\kappa=0.1$ a.u.) the problem becomes a standard,
linear $E\otimes e$ Jahn-Teller model, with adiabatic states $E_{\pm}(\boldsymbol{\rho})=\frac{1}{2}M\omega^{2}\rho^{2}\pm\kappa\rho$
(Fig. \ref{fig:Exact-quantum-dynamical results}), where $\rho$ is
the distance from the $z$ axis. We solved the time-dependent Schr\"{o}dinger
equation for this problem using a standard Split-Operator algorithm
in conjunction with Fast-Fourier-Transforms to go back and forth between
real- and momentum-space. The wavefunction was represented on a fine
grid (1024$\times$1024), which was centered around the CI point and
taken of length $20\,a_{0}$ along each direction. A small time step
of $\Delta t=0.1\,$a.u. was adopted to ensure a good sampling of
the geometric phase over time. The initial wavefunction was obtained
by combining a nuclear wavepacket $\psi_{0}(\mathbf{x})$ with a ground
electronic state $\ket{u_{-}(\mathbf{x})}$, i.e. $\braket{\mathbf{x}|\Psi_{0}}_{X}=\psi_{0}(\mathbf{x})\ket{u_{-}(\mathbf{x})}$.
The nuclear wavepacket $\psi_{0}(\mathbf{x})$ was a Gaussian centered
at $x_{0}=-2\kappa/M\omega^{2}$ and $y_{0}=0$, with ground-state
width along both $x$ and $y$ ($\Delta x=\Delta y=\sqrt{\hbar/2m\omega})$
and zero nominal momentum along both directions. As for the electronic
state, we fixed the \emph{gauge} with the choice $\ket{u_{-}(\mathbf{x})}=(-e^{-i\phi}\ket{1}+\ket{2})/\sqrt{2}$
where $\phi$ is the azimuthal angle of the position vector $\mathbf{x}$.
With these choices the initial wavefunction is current-carrying, and
presents a non-vanishing velocity field directed along the negative
$\phi$ direction, albeit very small (with the chosen parameters)
in the region where the wavepacket moves. A different choice of initial
state in described in the SM, which also provides details about the
two-state model, the numerical implementation of the dynamics and
the calculation of the geometric phase. The latter was performed at
every time step along a number of paths fixed in time, that were discretized
on the numerical grid. 

Fig. \ref{fig:Exact-quantum-dynamical results} shows the main results
of our numerical investigation. The wavepacket spreads along the valley
of the ``Mexican hat'' potential and its trailing edges meet each
other and interfere at time $t\approx75\,fs$, after which the wavepacket
covers more or less uniformly the valley, with a time-varying interference
pattern (panels (b)-(d)). Fig. \ref{fig:Exact-quantum-dynamical results}
(e) shows the evolution of the geometric phase (in $\pi$ units) along
three significant paths, i.e., three circles of radius $R=2.0,2.5$
and $3.0\,a_{0}$ centered at the CI point. The behaviour of the phase
at short time ($t\lessapprox20\,fs$) is somewhat uncertain because of
the intrinsic limits of the numerical implementation: the paths go
through regions in space where the nuclear density is very small and
the polarization field, as well as the momentum and velocity fields,
get affected by sizable numerical errors (see SM for a discussion
on this point and our fix). This problem, however, has no physical
implications since the geometric phase is of little practical utility
when the probe path along which the phase is computed lies entirely
(or partially) in a region where the system has little probability
to be found. After this transient, the phase is seen to undergo an
evident variation \textemdash{} which is numerically robust \textemdash{}
when the wavepacket edges start to interfere, and later gets back
to the value expected for an adiabatic dynamics (Fig. \ref{fig:Exact-quantum-dynamical results}
(e)). This provides direct evidence for the transition between geometric
and topological phase, from the vantage point of exact quantum dynamics.
Analysis of the contributions to the electromotive force confirms
that the NBO component contributes little to the phase change, while
the \emph{pseudo}-electric one plays the major role (see SM).

In this example the dynamics remains adiabatic to a large extent,
with the excited-state probability never exceeding $10^{-5}$. Non-adiabatic
behaviour appears as a transient, and it is more marked closer to
the CI, in accordance with the stronger non-adiabatic coupling experienced
by the wavepacket there. In this situation, the concept of adiabatic
geometric phase remains highly relevant, when the phase itself is
computed along physically meaningful paths. 

\textbf{\emph{Conclusions.}} We have shown that the molecular geometric
phase of the adiabatic approximation can be seamlessly extended to
exact quantum dynamics. The generalized phase is time dependent and
its evolution is governed by a (reversed) Maxwell-Faraday induction
law, with non-conservative forces arising from the electron dynamics
that play the role of electromotive forces. Though generally evolving
in a complicated way, this geometric phase remains highly relevant
when the dynamics is close to adiabatic and a physically motivated
choice of the path is performed. 


%
\pagebreak{}

\section*{Supplemental Material}

\section{Dynamical\emph{ }Paths}

We consider here the total rate of change of the phase of Eq. \ref{eq:definition of QHD phase},
evaluated along a generic curve $\gamma$ that connects $\mathbf{x}_{a}$
to $\mathbf{x}_{b}$ at initial time and that follows the fluid flow,
i.e. $\gamma(t):[0,1]\ni s\rightarrow\mathbf{x}(s;t)$ for some convenient
parameterization of the curve, $\mathbf{x}(0;t_{0})=\mathbf{x}_{a}$,
$\mathbf{x}(1;t_{0})=\mathbf{x}_{b}$ and $\partial\mathbf{x}(s,t)/\partial t=\mathbf{v}(\mathbf{x}(s,t),t)$,
where $v^{k}(\mathbf{x},t)$ is the velocity field of the probability
fluid at time $t$. We denote this phase $\tilde{\Gamma}[\gamma]$
to emphasize that it is related to a path that is tied to the fluid
dynamics, i.e.,
\[
\tilde{\Gamma}[\gamma](t):=-\frac{1}{\hbar}\sum_{k}\int_{\gamma(t)}dx^{k}\pi_{k}(t)
\]
where $\pi_{k}(t)$ is the particle momentum field at time $t$, to
be evaluated along $\gamma$. The case of a closed curve, $\tilde{\Gamma}_{O}[\gamma]$,
then follows from the results given below upon setting $\mathbf{x}_{b}(t)=\mathbf{x}_{a}(t)$,
where $\mathbf{x}_{a}(t)$ ($\mathbf{x}_{b}(t)$) is the position
at time $t$ of a particle that was in $\mathbf{x}_{a}$ ($\mathbf{x}_{b}$)
at initial time $t_{0}$. In contrast to the case first addressed
in the manuscript (where the path was fixed in time) the rate of change
of the line integral involves the \emph{material} derivative of the
momentum and contains contributions from the moving path
\[
-\hbar\frac{d\tilde{\Gamma}[\gamma]}{dt}=\sum_{k}\int_{\gamma}dx^{k}\dot{\pi}_{k}+\sum_{kj}\int_{\gamma}dx^{k}\pi_{j}\partial_{k}v^{j}
\]
Here, the first term involves the \emph{total} force acting on the
fluid elements (Eq. \ref{eq:momentum equation})
\[
F_{k}^{\text{tot}}=-\text{Tr}_{\text{el}}(\rho_{\text{el}}\partial_{k}H_{\text{el}})-\frac{\hbar^{2}}{n}\sum_{ij}\xi^{ij}\partial_{i}(ng_{kj})-\partial_{k}Q
\]
while for the second term we have
\[
\sum_{j}\pi_{j}\partial_{k}v^{j}=\sum_{ij}\xi^{ij}\pi_{j}(\partial_{k}\pi_{i})\equiv\partial_{k}T
\]
where $T=\sum_{ij}\xi^{ij}\pi_{i}\pi_{j}$ is the classical kinetic
energy. As a result, 
\[
-\hbar\frac{d\tilde{\Gamma}[\gamma]}{dt}=\sum_{k}\int_{\gamma}dx^{k}F_{k}^{\text{tot}}+\Delta T
\]
where $\Delta T=T(\mathbf{x}_{b}(t),t)-T(\mathbf{x}_{a}(t),t)$. For
a closed path the second term disappears and the force entering the
first line integral can be simplified to its classical-like contribution
\[
F_{k}=-\text{Tr}_{\text{el}}(\rho_{\text{el}}\partial_{k}H_{\text{el}})-\frac{\hbar^{2}}{n}\sum_{ij}\xi^{ij}\partial_{i}(ng_{kj})
\]
The net result for the rate of change of the phase is thus Eq. \ref{eq:rate of phase change - dynamical loop}
of the main text, 
\[
-\frac{d\tilde{\Gamma}_{O}[\gamma]}{dt}=\mathfrak{E}^{\text{NBO}}+\mathfrak{E}^{\text{el}}
\]
This shows that the magnetic contribution $\mathfrak{E}^{\text{mag}}$
appearing in Eq. \ref{eq:rate of phase change - fixed loop} is the
phase transported along the flow per unit time, i.e. a drift term
that only appears when the loop is held fixed in time.

\section{Two-state\emph{ }problem}

\subsection{Generalities}

In the two-state approximation the electron-nuclear dynamical problem
is governed by the Hamiltonian
\[
H=\frac{1}{2}\sum_{ij}\xi^{ij}\hat{p}_{i}\hat{p}_{j}+H_{\text{el}}(\mathbf{x})
\]
where the first term is the nuclear kinetic energy and $H_{\text{el}}(\mathbf{x})$,
the electronic Hamiltonian, takes the form
\[
H_{\text{el}}(\mathbf{x})=A(\mathbf{x})\mathbb{I}_{2}+\mathbf{B}(\mathbf{x})\boldsymbol{\sigma}
\]
in the spinor representation of the wavefunction 
\[
\boldsymbol{\Psi}(\mathbf{x})=\left(\begin{array}{c}
\Psi_{1}(\mathbf{x})\\
\Psi_{2}(\mathbf{x})
\end{array}\right)\in L^{2}(\mathcal{M})\otimes\mathbb{C}^{2}
\]
where $\Psi_{\sigma}(\mathbf{x})$ gives the probability amplitudes
of finding the system in the $\sigma^{\text{th}}$ diabatic state.
Here, $\mathbf{B}(\mathbf{x})\in\mathcal{N}\cong\mathbb{R}^{3}$ is
an effective magnetic field for the pseudo-spin describing the local
electronic state, whose magnitude $B$ determines the size of the
energy gap $E_{\text{gap}}(\mathbf{x})=2B(\mathbf{x})$ and whose
direction $\mathbf{b}(\mathbf{x})$ specifies the local ``principal
axis'' of the electronic system. Both $B$ and $\mathbf{b}$ can
depend on $\mathbf{x}$, thereby determining complex (adiabatic) energy
landscapes, $E_{\pm}(\mathbf{x})=A(\mathbf{x})\pm B(\mathbf{x})$,
and non-trivial (i.e., coordinate-dependent) diabatic-to-adiabatic
transformations (local rotations of the frame) \citep{Kouppel1984}.
Notice that $\mathbf{B}$ affects the geometric properties of the
adiabatic bundles while $A(\mathbf{x})$ ``tunes'' the nuclear dynamics.

In the QHD-EF framework the local electronic state is described by
a conditional density matrix, $\rho_{\text{el}}(\mathbf{x})=\frac{1}{2}(\sigma_{0}+\mathbf{s}(\mathbf{x})\boldsymbol{\sigma})$,
where $\mathbf{s}(\mathbf{x})\in\mathcal{N}$ is the polarization
vector, with $\lVert\mathbf{s}\rVert=1$ since $\rho_{\text{el}}$
describes a pure state. The electronic equation of motion, Eq. \ref{eq:LvN equation for electrons}
with the \emph{e-n} coupling term given by Eq. \ref{eq:e-n coupling Hamiltonian},
can thus be recast as an equation of motion for $\mathbf{s}\equiv\mathbf{s}(\mathbf{x},t)$.
Eq. \ref{eq:LvN equation for electrons} gives
\[
i\hbar\dot{\mathbf{s}}\boldsymbol{\sigma}=[\mathbf{B}\boldsymbol{\sigma}+\delta H_{\text{en}},\mathbf{s}\boldsymbol{\sigma}]
\]
where the dot stands for the material derivative and
\[
\delta H_{\text{en}}=-\frac{\hbar^{2}}{4n}\sum_{ij}\xi^{ij}\partial_{i}(n\mathbf{s}_{j}\boldsymbol{\sigma})
\]
with $\mathbf{s}_{j}\equiv\partial_{j}\mathbf{s}$. Here, upon repeated
use of the identity $(\mathbf{a}\boldsymbol{\sigma})(\mathbf{b}\boldsymbol{\sigma})=\mathbf{a}\mathbf{b}+i(\mathbf{a}\times\mathbf{b})\boldsymbol{\sigma}$,
we find $[\mathbf{B}\boldsymbol{\sigma},\mathbf{s}\boldsymbol{\sigma}]=2i(\mathbf{B}\times\mathbf{s})\boldsymbol{\sigma}$
and $[\partial_{i}(n\mathbf{s}_{j}\boldsymbol{\sigma}),\mathbf{s}\boldsymbol{\sigma}]=2i(\partial_{i}n)(\mathbf{s}_{j}\times\mathbf{s})\boldsymbol{\sigma}+2in\,(\partial_{i}\mathbf{s}_{j}\times\mathbf{s})\boldsymbol{\sigma}$,
hence
\[
\hbar\dot{\mathbf{s}}\boldsymbol{\sigma}=2(\mathbf{B}\times\mathbf{s})\boldsymbol{\sigma}-\frac{\hbar^{2}}{2n}\sum_{ij}\xi^{ij}\{\partial_{i}n(\mathbf{s}_{j}\times\mathbf{s})\boldsymbol{\sigma}+n(\partial_{i}\mathbf{s}_{j}\times\mathbf{s})\boldsymbol{\sigma}\}
\]
Multiplying by $\boldsymbol{\sigma}$ and tracing over the pseudo-spin
degrees of freedom we arrive at the desired equation of motion for
$\mathbf{s}$
\begin{align*}
\dot{\mathbf{s}} & =\frac{2}{\hbar}B(\mathbf{b}\times\mathbf{s})-\frac{\hbar}{2n}\sum_{ij}\xi^{ij}\{\partial_{i}n(\mathbf{s}_{j}\times\mathbf{s})+n(\partial_{i}\mathbf{s}_{j}\times\mathbf{s})\}\\
 & =(\Omega\mathbf{b}+\boldsymbol{\tau})\times\mathbf{s}-\frac{\hbar}{2}\sum_{ij}\xi^{ij}\partial_{i}(\mathbf{s}_{j}\times\mathbf{s})
\end{align*}
which is Eq. \ref{eq:2-state electron dynamics} of the main text
with $\Omega=2B/\hbar$ and $\boldsymbol{\tau}=-\hbar/2\sum_{ij}\xi^{ij}\partial_{i}\ln n\,\mathbf{s}_{j}\equiv\sum_{ij}\xi^{ij}w_{i}\mathbf{s}_{j}=\sum_{j}u^{j}\mathbf{s}_{j}$.
Here, $w_{k},u^{k}$ are the $k^{\text{th}}$ components of the ``osmotic''
momentum and velocity, respectively, defined as the imaginary parts
of the complex-valued momentum and velocity fields, $\Pi_{k}=\hat{\pi}_{k}\psi/\psi$
and $V^{k}=\hat{v}^{k}\psi/\psi$ where $\psi$ is the nuclear wavefunction
and $\hat{\pi}_{k}$, $\hat{v}^{k}$ are the Schr\"{o}dinger-representation
mechanical momentum and velocity operators. The equation of motion
for $\mathbf{s}$ is purity-conserving, i.e. $\mathbf{s}\dot{\mathbf{s}}=0$,
as must be the case since the original EF electronic equation was
for a pure local electronic state.

As for the nuclear dynamics, the Ehrenfest force is readily expressed
in terms of the polarization vector
\[
F_{k}^{\text{Eh}}=-\text{Tr}_{\text{el}}(\rho_{\text{el}}\partial_{k}H_{\text{el}})=-(\partial_{k}A+\mathbf{s}\partial_{k}\mathbf{B})
\]
while the \emph{pseudo}-electric contribution requires that the FS
(pseudo)metric $g_{kj}$ is expressed in terms of $\mathbf{s}$. We
proceed with calculating the quantum geometric tensor $q_{kj}$, since
this also provides the Berry curvature of the underlying electronic
state. This is simple when using Eq. \ref{eq:quantum geometric tensor},
with $\rho_{\text{el}}$ given in terms of $\mathbf{s}$. The result
is Eq. \ref{eq:2-state quantum geometric tensor} of the main text
\[
q_{kj}=\frac{1}{4}\left(\mathbf{s}_{k}\mathbf{s}_{j}+i\mathbf{s}(\mathbf{s}_{k}\times\mathbf{s}_{j})\right)
\]
which gives the metric ($g$) and curvature ($\mathcal{B}$) tensors
in the illuminating forms
\[
g=\frac{1}{4}d\mathbf{s}\,d\mathbf{s}\ \ \ \mathcal{B}=-\frac{\hbar}{2}\mathbf{s}(d\mathbf{s}\times d\mathbf{s})
\]
where $d\mathbf{s}$ is a three-component 1-form, arranged as a vector
of $\mathcal{N}$. To avoid confusion, their actions on arbitrary
tangent vectors $\mathbf{u},\mathbf{v}\in T_{x}\mathcal{M}$ read
as
\[
g(\mathbf{u},\mathbf{v})=\frac{1}{4}(\partial_{\mathbf{u}}\mathbf{s})\,(\partial_{\mathbf{v}}\mathbf{s})\ \ \ \mathcal{B}(\mathbf{u},\mathbf{v})=-\frac{\hbar}{2}\mathbf{s}\left((\partial_{\mathbf{u}}\mathbf{s})\times(\partial_{\mathbf{v}}\mathbf{s})\right)
\]
where $\partial_{\mathbf{u}}$ stands for the directional derivative
along $\mathbf{u}$, $\partial_{\mathbf{u}}f=\sum_{j}u^{j}\partial_{j}f$.
Here, the vectors $\partial_{\mathbf{u}}\mathbf{s}$ are all tangent
to the Bloch sphere $S^{2}\subset\mathcal{N}$ swept by the polarization
vectors $\mathbf{s}$ (i.e., the projective Hilbert space of the electronic
system), since $\mathbf{s}\mathbf{s}_{j}=0$ is required by the pure-state
nature of the electronic state. Note that $g$ is generally a \emph{pseudo}
metric in this context, in particular in this 2-state problem where
the state space is the two-dimensional real manifold $S^{2}$, and
the $\partial_{j}\mathbf{s}$ 's are necessarily linearly dependent
on each other when the number $N$ of nuclear degrees of freedom exceeds
two. 

The geometric properties of the adiabatic bundles can be obtained
from the above expression of the quantum geometric tensor upon setting
$\mathbf{s}(\mathbf{x})=\pm\mathbf{b}(\mathbf{x})$ for the upper
and lower energy state, respectively, and depend on the details of
the given electronic Hamiltonian, more precisely on the function $\beta:\mathcal{M}\ni\mathbf{x}\rightarrow\mathbf{B}(\mathbf{x})\in\mathcal{N}$
specifying the magnetic field (i.e. how its orientation varies in
nuclear configuration space). However, they can be ``pulled back''
from those of the spin-problem in a magnetic field, where $\mathcal{N}$
is the parameter space and the cartesian components of the magnetic
field form a set of convenient coordinates \citep{Berry1984}. In
other words, the interesting geometric tensor for our problem, $q$,
is the pull-back of the geometric tensor $\tilde{q}$ (a section of
$T^{*}\mathcal{N}\times T^{*}\mathcal{N}$) describing the adiabatic
dynamics of a spin in a slowing varying magnetic field, i.e., $q_{\mathbf{x}}(\mathbf{u},\mathbf{v})=(\beta^{*}\tilde{q})_{\mathbf{x}}(\mathbf{u},\mathbf{v})\equiv\tilde{q}_{\beta(\mathbf{x})}(d\beta_{\mathbf{x}}(\mathbf{u}),d\beta_{\mathbf{x}}(\mathbf{v}))$
for $\mathbf{x}\in\mathcal{M}$ and $\mathbf{u},\mathbf{v}\in T_{\mathbf{x}}\mathcal{M}$.
The components of the metric tensor $\tilde{g}=\Re\tilde{q}$ read
(for both the upper and the lower energy state) as
\begin{align*}
\tilde{g}_{kj} & =\frac{1}{4}\mathbf{b}_{k}\mathbf{b}_{j}=\frac{1}{4B^{2}}\left(\delta_{kj}-\frac{B_{k}B_{j}}{B^{2}}\right)\\
 & =\frac{1}{4B^{2}}\left(\frac{2}{3}\delta_{kj}-\frac{Q_{kj}}{B^{2}}\right)
\end{align*}
where $Q$ is the traceless tensor defined by $Q_{kj}=3B_{k}B_{k}-\delta_{kj}B^{2}$.
This follows from $\mathbf{b}_{k}=\partial_{k}(\mathbf{B}/B)=\frac{1}{B}\left(\mathbf{e}_{k}-\frac{B_{k}}{B}\mathbf{b}\right)$
and gives the required tensor in the form $g=\sum_{kj}\tilde{g}_{kj}dB^{k}dB^{j}$
where $dB^{k}\equiv\sum_{l}(\partial_{l}B^{k})dx^{l}$, for the specific
magnetic function $\beta$ defining the electronic problem. Note that
$\tilde{g}$ above is a \emph{pseudo}-metric and reduces to the proper
Fubini-Study metric of $S^{2}$ when $\mathbf{B}$ is constrained
to have a fixed magnitude (as the latter does not affect the spin
state). As for the curvature, its components on the canonical basis
$dB^{k}dB^{j}$ turn out to be
\begin{align*}
B_{kj}^{\pm} & =\mp\frac{\hbar}{2B^{2}}\mathbf{b}\left(\mathbf{e}_{k}-\frac{B_{k}}{B}\mathbf{b}\right)\times\left(\mathbf{e}_{j}-\frac{B_{j}}{B}\mathbf{b}\right)\\
 & \equiv\mp\frac{\hbar}{2B^{3}}\mathbf{B}(\mathbf{e}_{k}\times\mathbf{e}_{j})
\end{align*}
and give the curvature tensor in the form
\[
\tilde{\mathcal{B}}_{\pm}=H_{x}^{\pm}dB^{y}\wedge dB^{z}+H_{y}^{\pm}dB^{z}\wedge dB^{x}+H_{z}^{\pm}dB^{x}\wedge dB^{y}
\]
where $\wedge$ denotes the wedge product and 
\[
\mathbf{H}^{\pm}=(H_{x}^{\pm},H_{y}^{\pm},H_{z}^{\pm})^{t}=\mp\frac{\hbar}{2}\frac{\mathbf{B}}{B^{3}}
\]
is the ``Berry field'' describing a magnetic monopole of charge
$q_{\pm}=\mp\hbar/2$ at the origin of $\mathcal{N}$. The curvature
tensor for the given electronic problem is thus obtained from the
above expression for $\tilde{\mathcal{B}}_{\pm}$ upon using $dB^{i}\wedge dB^{j}=\sum_{kl}J_{kl}^{ij}\,dx^{k}\wedge dx^{l}$,
where $J_{kl}^{ij}=\text{det}[\partial(B^{i},B^{j})/\partial(x^{k}x^{l})]$
is a Jacobian. 

The above results allow one to compute the Berry phase along arbitrary
closed paths $\gamma\subset\mathcal{M}$ in parameter space through
its mapping to $\mathcal{N}$ space, i.e. by considering the image
$\tilde{\gamma}=\beta\circ\gamma$ generated by the magnetic field
function $\beta$ (the ``gap function''). As is well known \citep{Berry1984,Bohm2003},
the Berry phase along a loop $\tilde{\gamma}\subset\mathcal{N}$ is
geometric and is given by $q_{\pm}$ times the solid angle subtended
by loop. This is the flux of the Berry field $\mathbf{H}^{\pm}$ through
an arbitrary surface subtending $\tilde{\gamma}$ (in $\hbar$ units,
with our convention). In a typical molecular problem, however, one
of the $\mathbf{B}$ components identically vanishes because of time-reversal
symmetry, say $B_{z}$. Hence, $H_{z}=0$, $dB^{z}\equiv0$ and $\mathcal{B}_{\pm}$
vanishes identically except at the CI seam. In this case, the image
paths $\tilde{\gamma}$ necessarily lie on the $xy$ plane of the
$\mathcal{N}$ space and the Berry phase becomes $\mp n\pi$, where
$n$ is the winding number of $\tilde{\gamma}$ around the origin
of $\mathcal{N}$. 

\subsection{Model problem}

For concreteness we consider here a model problem with 3 nuclear degrees
of freedom ($\mathcal{M}\cong\mathbb{R}^{3})$ and a diagonal mass
tensor, $\xi^{ij}=\delta^{ij}M^{-1}$. The problem is defined by $\mathbf{B}=Q\boldsymbol{\rho}$
\textemdash{} where $\boldsymbol{\rho}$ is the projection of the
position vector on the \emph{xy} plane and $Q$ a real constant \textemdash{}
and exemplifies the situation of a CI seam (the $z$ axis) that presents,
as mentioned above, a topological phase. This situation closely resembles
the common problem of a spin in an arbitrarily varying magnetic field,
except for an important topological difference. The latter problem
is indeed defined by $\mathbf{B}=\mathbf{r}$ (henceforth, $\mathbf{r}$
is the position vector in $\mathcal{N}\cong\mathbb{R}^{3}$) and features
an isolated CI point (the origin of $\mathbb{R}^{3}$) rather than
a CI seam. As a consequence, the parameter manifold for the adiabatic
dynamics is simply connected and cannot display any topological phase
other than the trivial one, in contrast to the problem considered
here where the adiabatic manifold is multiply connected and a topological
phase appears naturally when its curvature vanishes. 

The nuclear dynamics is also determined by the function $A=A(\mathbf{x})$
that ``tunes'' the adiabatic potentials and that, depending on its
strength, can be used to control the wavepacket dynamics. This function
is irrelevant, though, for the geometric properties of interest here
and in the numerical applications discussed below will be taken independent
of $z$. This helps reducing the dynamical problem to two nuclear
degrees of freedom and two coupled electronic states.

\subsubsection{Initialization of the dynamics}

We are interested in the dynamics of the system when it is initially
prepared in an adiabatic state, i.e., when the initial wavefunction
takes the form
\begin{equation}
\ket{\Psi_{n}}=\int_{X}d\mathbf{x}\psi(\mathbf{x})\ket{u_{n}(\mathbf{x})}\ket{\mathbf{x}}\label{eq:true adiabatic state}
\end{equation}
where $n=\pm1$, $\ket{u_{n}}=\chi_{n}^{1}\ket{1}+\chi_{n}^{2}\ket{2}$,
and $\chi_{n}(\mathbf{x})$ is a (normalized) spinor with definite
projection on the local $\mathbf{B}$ axis \textemdash{} i.e., such
that $(\mathbf{b}\boldsymbol{\sigma})\chi_{\pm}=\pm\chi_{\pm}$. The
latter represents the adiabatic electronic state in the diabatic basis
$\{\ket{1},\ket{2}\}$. 

We begin by addressing certain subtleties inherent in the problem
of a spin in a magnetic field, focusing for definiteness on the upper
energy state (similar results follow for the lower one). We use the
standard parametrization 
\begin{equation}
\ket{u_{+}}=\cos(\theta/2)\ket{1}+e^{i\phi}\sin(\theta/2)\ket{2}\label{eq:northern gauge}
\end{equation}
for the spin state with +1 projection on the direction $\mathbf{n}=\mathbf{n}(\theta,\phi)$
identified by the polar ($\theta$) and the azimuthal ($\phi$) angles
of the usual spherical coordinates. As is well known, this parametrization
is regular everywhere on $\mathcal{S}^{2}$ except at the south pole
where $\ket{u_{+}}$ is seen to give $\ket{2}$ with a phase factor
that depends on the way the south pole is approached (that is, on
the irrelevant angle $\phi$). Therefore a second parametrization
is required, e.g., 
\begin{equation}
\ket{\tilde{u}_{+}}=e^{-i\phi}\cos(\theta/2)\ket{1}+\sin(\theta/2)\ket{2}\equiv e^{-i\phi}\ket{u_{+}}\label{eq:southern gauge}
\end{equation}
which is related to the previous one by a\emph{ gauge} transformation.
The \emph{gauge} transformation involved here, however, is \emph{non-standard}:\emph{
}it is a ``radical'' transformation defined by the function $\varphi(\phi)=\phi$
which increases by $2\pi$ after making a turn around the $z$ axis,
rather than getting back to its original value. This is a necessary
feature of the transformation if $\ket{\tilde{u}_{+}}$ has to fix
the problems of $\ket{u_{+}}$, since these problems are unaffected
by phase transformation functions $\varphi$ which are smooth all
over the manifold (as the ones considered in the main text). These
two parametrizations are sufficient to introduce proper frames in
the whole adiabatic bundle $\pi:\mathcal{E}_{+}\rightarrow S^{2}$,
i.e. the vector bundle based on $S^{2}$ whose fibers $\pi^{-1}\{\mathbf{s}\}$
are the $+$ eigenspaces of $\mathbf{s}\boldsymbol{{\sigma}}$. 

Now, since in the spin problem the full set of states (i.e. points
of $S^{2}$) is required to define a global adiabatic state, a globally
smooth electronic state (hence a globally smooth nuclear wavefunction)
\emph{cannot} be introduced. Rather, in the adiabatic wavefunction
for the total system one has to use a pair $\{\psi_{0},\ket{u_{+}}\}$
for $\mathbf{r}$ such that $\mathbf{b}(\mathbf{r})\neq-\mathbf{e}_{3}$
and a second pair $\{\tilde{\psi}_{0},\ket{\tilde{u}_{+}}\}$ for
$\mathbf{b}(\mathbf{r})\neq+\mathbf{e}_{3}$, with $\tilde{\psi}_{0}=\psi_{0}e^{i\phi}$
in overlapping regions. Here, $\psi_{0}(\mathbf{r})$ and $\tilde{\psi}_{0}(\mathbf{r})$
are nuclear wavefunctions representing the same state in the two different
\emph{gauges} defined above and the $\mathbf{e}_{k}$'s are canonical
orthonormal vectors in $\mathcal{N}\cong\mathbb{R}^{3}$. In this
problem the field direction is also the direction of the position
vector, $\mathbf{b}\equiv\hat{\mathbf{r}}$, and the spherical coordinates
$(\theta,\phi)$ become the spherical angles of the nuclear position
vector $\mathbf{r}$. Hence, one can use $\psi_{0}(\mathbf{r})$ and
$\ket{u_{+}(\mathbf{\hat{r}})}$ for the whole space except a small
cone around the negative $z$ axis, or the pair $\tilde{\psi}_{0}(\mathbf{r})$
and $\ket{\tilde{u}_{+}(\mathbf{\hat{r}})}$ everywhere except a small
cone around the positive real axis. In other words, $\tilde{\psi}(\mathbf{r})=e^{i\phi}\psi(\mathbf{r})$
holds everywhere except along $z$, where $\phi$ would be undefined
in any case. 

The vector potentials corresponding to the above \emph{gauge} choices
can be computed with a direct calculation and turn out to be, respectively,
\[
\mathbf{A}=-\frac{1}{2}\frac{\tan(\theta/2)}{r}\mathbf{e}_{\phi}\ \ \tilde{\mathbf{A}}=-\frac{1}{2}\frac{\cot(\theta/2)}{r}\mathbf{e}_{\phi}
\]
where $\mathbf{e}_{\phi}$ is the unit vector along $\phi$. They
clearly present a singularity, respectively, for $\theta=\pi$ and
$\theta=0$ (the Dirac strings), but both describe the same Berry
field where they apply, i.e. $\hbar\boldsymbol{\nabla}\times\mathbf{A}=\hbar\boldsymbol{\nabla}\times\tilde{\mathbf{A}}=\mathbf{H}^{+}$.
When the underlying electronic states are paired with suitable nuclear
wavefunctions, these vector potentials determine the mechanical momentum
and velocity fields, and fix their circulation, \emph{modulo an appropriate
``quantum''}. In a sense they fix the symmetry of the momentum and
velocity fields and their vortex structure. Notice though that there
is nothing special about the $z$ axis. Suitable\emph{ gauge} transformations
exist that reorient the $z$ axis and the ensuing electronic frame
can be used to align the symmetry axis of the nuclear velocity field
(and its half vortex-line) along \emph{arbitrary} directions.

For definiteness we choose the first \emph{gauge} and first set $\psi$
to be real. The momentum field then reads as 
\[
\boldsymbol{\pi}=+\frac{\hbar}{2}\frac{\tan(\theta/2)}{r}\mathbf{e}_{\phi}
\]
everywhere except on the negative $z$ axis (and where $\psi_{0}$
eventually vanishes). Like $\mathbf{A}$, this field has a singularity
on the negative $z$ axis and this cannot be removed by changing the
\emph{gauge} with a regular transformation, since $\boldsymbol{\pi}$
is \emph{gauge} invariant and the above expression holds for arbitrarily
small cones around the negative $z$ axis, i.e. it cannot be remedied
with any choice of $\tilde{\psi}_{0}$ on the $z$ axis. And choosing
$\psi_{0}$ with a node on the negative $z$ axis removes the problem
of using a pair of \emph{gauges} (since $\ket{u_{+}}$ for $\theta=\pi$
is then useless) but does not remove the singularity in the momentum
and velocity fields. The circulation of this momentum field along
a circular path at constant latitude (as defined by Eq. \ref{eq:definition of QHD phase})
gives the well-known adiabatic result
\[
\Gamma_{O}(\theta)=-\frac{1}{2}\int_{0}^{2\pi}\frac{\tan(\theta/2)}{r}r\sin(\theta)d\phi=-2\pi\sin^{2}\left(\frac{\theta}{2}\right)
\]
which represents the flux of the Berry field on the sphere portion
which is above $\theta$. For $\theta$ tending to $\pi$, i.e. when
the loop shrinks to a point on the \emph{negative} $z$ axis, this
tends to $-2\pi$, which is yet zero but only mod $2\pi$. This amounts
to a non-vanishing first Chern number, here of value -1, that reflects
the non-trivial topology of $\mathcal{E}_{+}$ and that manifests
itself in a vortex structure of the momentum field of the $+$ adiabatic
state on the negative $z$ axis, with the above \emph{gauge} choice.
Replacing $\psi_{0}(\mathbf{r})$ with an arbitrary but smooth function
changes the momentum field locally but not its circulation, hence
the same result holds for \emph{arbitrary} nuclear wavefunctions.
In this sense, the \emph{gauge} choice determines the symmetry of
the momentum field and its vortex structure, modulo a longitudinal
vector contribution ($\hbar\boldsymbol{\nabla}\Im\ln\psi_{0}$). Clearly,
upon using $\ket{\tilde{u}_{+}}$ one defines a different class of
states, with a different circulation around that axis. The difference
as compared with the previous case is $2\pi$ and now a vortex structure
appears on the positive $z$ axis. And, as mentioned above, with an
appropriate rotation one can make the half-line vortex pointing in
arbitrary directions. Actually, this half-line vortex is not even
a necessary feature of the adiabatic state: the implicit assumption
above was that $\boldsymbol{\pi}$ was well-defined wherever $\mathbf{A}$
was not singular, i.e., that $\psi_{0}$ did not vanish somewhere
(otherwise a singularity would also arise from the canonical momentum
contribution, $\hbar\boldsymbol{\nabla}\Im\ln\psi_{0}$). If $\psi_{0}(\mathbf{r})$
has zeros one can exploit their presence to smoothly (but radically)
change the underlying \emph{gauge}. For instance, if $\psi_{0}(\mathbf{r})\equiv0$
in the $xy$ plane one can smoothly switch from $\ket{u_{+}}$ to
$\ket{\tilde{u}_{+}}$ and thus remove the half-line vortex. The price
to be paid is that now the singularity in the momentum field extends
to the $xy$ plane. In this case, the circulation $\Gamma_{O}(\theta)$
for the loops at constant latitude considered above suddenly jumps
from $-\pi$ to $\pi$ when $\theta=\pi/2$ and tends to zero for
both $\theta\rightarrow0$ and $\theta\rightarrow\pi$, being completely
undefined for $\theta=\pi/2$. 

The above ``frustration'' problems do not arise in our model molecular
problem, where the adiabatic state involves only the points at the
equator of $S^{2}$. In this case a single parametrization suffices,
e.g.
\begin{equation}
\ket{u_{+}}=\frac{1}{\sqrt{2}}(\ket{1}+e^{i\phi}\ket{2})\label{eq:northern gauge for JT}
\end{equation}
whose corresponding Berry vector potential reads as 
\[
\mathbf{A}=-\frac{1}{2\rho}\mathbf{e}_{\phi}
\]
Here, $\phi$ can be identified with the azimuthal angle of the position
vector and $\rho$ is the distance from the $z$ axis. As anticipated
above, the Berry field vanishes everywhere except on the $z$ axis
(i.e., $\boldsymbol{\nabla}\times\mathbf{A}\equiv\mathbf{0}$ where
$\mathbf{A}$ is well defined), however, the circulation of $\mathbf{A}$
around that axis is non-vanishing and describes a topological phase.
The latter can be obtained by considering a circular path of radius
$\rho$ (since this value is the same for any homotopic loop thanks
to $\boldsymbol{\nabla}\times\mathbf{A}=\mathbf{0}$), and it is easily
seen to be non-trivial, 
\[
\oint\mathbf{A}d\mathbf{x}=-\int_{0}^{2\pi}\frac{1}{2\rho}\rho d\phi=-\pi
\]
This is also the value (modulo $2\pi$) of the circulation of the
momentum field according to Eq. \ref{eq:definition of QHD phase},
with the above \emph{gauge} choice of the electronic wavefunction.
Indeed, arguing as above, if we take $\psi_{0}$ real and use the
above \emph{gauge} we have $\boldsymbol{\pi}=\frac{\hbar}{2\rho}\mathbf{e}_{\phi}$,
which gives $-\pi$ when Eq. \ref{eq:definition of QHD phase} is
evaluated along a loop encircling the $z$ axis once. And the same
value of the phase results when replacing $\psi_{0}(\mathbf{x})$
with an arbitrary smooth function, not vanishing in extended regions.
A radically different \emph{gauge} choice, e.g., 
\[
\ket{\tilde{u}_{+}}=\frac{1}{\sqrt{2}}(e^{-i\phi}\ket{1}+\ket{2})
\]
gives a different circulation (+$\pi$) when paired with smooth nuclear
wavefunctions. And, as above, one can even interpolate between the
two situations by selecting $\psi_{0}$ with a node in the $xy$ plane
and switching between the two \emph{gauges} when going from $z>0$
to $z<0$. 

In other words, when selecting an \emph{adiabatic} state, the circulation
of the momentum field (according to Eq. \ref{eq:definition of QHD phase})
is ``pinned'', by construction, to the adiabatic value, but variations
of $2\pi n$ are possible depending on the singularities of $\boldsymbol{\pi}$
due to either $\mathbf{A}$ (through its topological frustration)
or $\psi$ (through its nodal structure).

\subsubsection{Phase dynamics}

We now address the behavior of the phase defined by Eq. \ref{eq:definition of QHD phase},
when the dynamics is started from an adiabatic state. The electron
dynamics is described by the equation of motion of the polarization
vector which, as seen above, reads as
\[
\dot{\mathbf{s}}=(\Omega\mathbf{b}+\boldsymbol{\tau})\times\mathbf{s}-\frac{\hbar}{2M}\sum_{j}(\partial_{j}\mathbf{s}_{j})\times\mathbf{s}
\]
and the rate of phase change appearing on the r.h.s. of Eq. \ref{eq:rate of phase change - fixed loop}
can be expressed as 
\[
-\frac{d\Gamma_{O}[\gamma]}{dt}=\sum_{\text{X}}\oint_{\gamma}\Phi^{\text{X}}\ \ \text{X=NBO,el and mag}
\]
where $\Phi^{X}$ are the following 1-forms, 
\begin{equation}
\Phi^{\text{NBO}}=\hbar^{-1}\mathbf{B}d\mathbf{s}\label{eq:s-expression NBO}
\end{equation}
\begin{equation}
\Phi^{\text{el}}=\frac{1}{2}\boldsymbol{\tau}d\mathbf{s}-\frac{\hbar}{4M}\sum_{j}\partial_{j}\mathbf{s}_{j}\,d\mathbf{s}\label{eq:s-expression EL}
\end{equation}
 and 
\begin{equation}
\Phi^{\text{mag}}=+\frac{1}{2}\ (\boldsymbol{\nu}\times\mathbf{s})d\mathbf{s}\label{eq:s-expression MAG}
\end{equation}
Here, $\Omega=2B/\hbar$ is the ``intrinsic'' precession frequency,
$\boldsymbol{\tau}=\frac{1}{M}\sum_{j}w_{j}\mathbf{s}_{j}$ is the
``nuclear torque'' and $\boldsymbol{\nu}=\frac{1}{M}\sum_{j}\pi_{j}\mathbf{s}_{j}$
is the ``drift velocity'', $\pi_{k}$ and $w_{k}$ being, as usual,
the real and imaginary parts of the complex-valued field $\Pi_{k}=-i\hbar\partial_{k}\ln\psi-\hbar A_{k}$.
The advantage of this formulation is that now the circulations $\mathfrak{E}^{\text{X}}=\oint_{\gamma}\Phi^{\text{X}}$
are mapped onto the Bloch sphere $S^{2}$: for $\mathbf{x}$ moving
along a curve $\gamma$, \textbf{$\mathbf{s}(\mathbf{x})\in S^{2}$}
traces a curve $\tilde{\gamma}$ on the sphere, $\boldsymbol{\tau},\boldsymbol{\nu}$
and $d\mathbf{s}$ are tangent to the sphere, while $\mathbf{B}$
can have both tangent and normal components, depending on the real-space
position $\mathbf{r}$. This gives immediately a simple way to express
the geometric phase and its rate of change at any time, once the polarization
field $\mathbf{s}$ is known. 

To see this, consider the ``northern'' \emph{gauge} in the $\mathcal{N}$
space where the Bloch sphere can be embedded, i.e., $\ket{u_{+}}$
of Eq. \ref{eq:northern gauge}. The corresponding vector potential
$\mathbf{A}=-\tan(\theta/2)\mathbf{e}_{\phi}/2r$ can be written in
terms of $\mathbf{s}$ by noticing that
\[
\tan(\theta/2)=\sqrt{\frac{1-\cos\theta}{1+\cos\theta}}=\sqrt{\frac{1-s_{z}}{1+s_{z}}}
\]
and 
\[
\mathbf{e}_{\phi}=-\frac{s_{y}}{\sqrt{s_{x}^{2}+s_{y}^{2}}}\mathbf{e}_{1}+\frac{s_{x}}{\sqrt{s_{x}^{2}+s_{y}^{2}}}\mathbf{e}_{2}
\]
where $s_{x},s_{y}$ and $s_{z}$ are the cartesian components of
$\mathbf{s}$. The dependence of $\mathbf{A}$ on $\mathbf{s}$ is
specific of this \emph{gauge} choice (since $\mathbf{A}$ is \emph{gauge}
dependent while $\mathbf{s}$ is not) but the circulation of $\mathbf{A}$
is the same for any \emph{gauge} choice (everywhere on $\mathcal{S}^{2}$
except at the south pole where $\mathbf{A}$ is ill-defined). Therefore,
for a curve $\gamma$ that is mapped onto the curve $\tilde{\gamma}$
on the sphere, the equation
\begin{equation}
\oint_{\tilde{\gamma}}\mathbf{A}d\mathbf{s}=-\frac{1}{2}\oint_{\tilde{\gamma}}\sqrt{\frac{1-s_{z}}{1+s_{z}}}\frac{(\mathbf{e}_{3}\times\mathbf{s})}{\sqrt{s_{x}^{2}+s_{y}^{2}}}d\mathbf{s}\label{eq:s-expression for Berry-phase}
\end{equation}
provides a convenient expression for the geometric phase in terms
of $\mathbf{s}$ only. This expression is valid for arbitrary curves,
provided their images $\tilde{\gamma}$ do not pass through the south
pole. This limitation can be easily overcome, however, since the position
of the south pole is arbitrary and can be changed with a rotation
of the coordinate system, i.e. one can always pick a point of $\mathcal{S}^{2}$
not belonging to $\tilde{\gamma}$ and use it to orient the south
pole when computing the phase using Eq. \ref{eq:s-expression for Berry-phase}.

Eq. \ref{eq:s-expression for Berry-phase} and Eqs. \ref{eq:s-expression NBO},
\ref{eq:s-expression EL} and \ref{eq:s-expression MAG} allow one
to extract the key dynamical information from an exact wavepacket
dynamics, without resorting to the exact factorization of the wavefunction.
Specifically, at any time, given the spinor field $\Psi_{\sigma}(\mathbf{x})$
representing the two-state \emph{e-n} wavefunction in the electronic
diabatic basis $\{\ket{1},\ket{2}\}$, the polarization field can
be written as $\mathbf{s}(\mathbf{x})=\boldsymbol{\Sigma}(\mathbf{x})/n(\mathbf{x})$
where $n(\mathbf{x})=\boldsymbol{\Psi}^{\dagger}(\mathbf{x})\boldsymbol{\Psi}(\mathbf{x})\equiv\sum_{\sigma}|\Psi_{\sigma}(\mathbf{x})|^{2}$
is the nuclear density and $\boldsymbol{\Sigma}(\mathbf{x})=\boldsymbol{\Psi}^{\dagger}(\mathbf{x})\boldsymbol{\sigma}\boldsymbol{\Psi}(\mathbf{x})$
or, explicitly, 
\begin{align*}
\boldsymbol{\Sigma}(\mathbf{x}) & =2\Re(\Psi_{1}^{*}(\mathbf{x})\Psi_{2}(\mathbf{x}))\mathbf{e}_{1}+\\
 & 2\Im(\Psi_{1}^{*}(\mathbf{x})\Psi_{2}(\mathbf{x}))\mathbf{e}_{2}+\\
 & (|\Psi_{1}(\mathbf{x})|^{2}-|\Psi_{2}(\mathbf{x})|^{2})\mathbf{e}_{3}
\end{align*}
The field $\mathbf{s}$ describes the local electronic states and
can be used to compute the geometric phase along arbitrary paths $\gamma:u\rightarrow\mathbf{x}(u)$,
through their images $\tilde{\gamma}:u\rightarrow\mathbf{s}(\mathbf{x}(u))$,
as well as the non-conservative fields $\mathbf{f}^{\text{X}}$ generating
the electromotive forces of Eqs. \ref{eq:s-expression NBO}, \ref{eq:s-expression EL}
and \ref{eq:s-expression MAG} (i.e. through $\Phi^{\text{X}}=\mathbf{f}^{\text{X}}d\mathbf{s}$).
The latter can be recast as 
\begin{equation}
\mathbf{f}^{\text{NBO}}=\hbar^{-1}\mathbf{B}\label{eq:NBO non-conservative field}
\end{equation}
\begin{equation}
\mathbf{f}^{\text{el}}=-\frac{1}{2Mn}\sum_{j}\left(w_{j}\partial_{j}\mathbf{\boldsymbol{\Sigma}}+\frac{\hbar}{2}\partial_{j}^{2}\mathbf{\boldsymbol{\Sigma}}\right)\label{eq:EL non-conservative field}
\end{equation}
\begin{equation}
\mathbf{f}^{\text{mag}}=-\frac{1}{2Mn}\left(\mathbf{s}\times\sum_{j}\pi_{j}\partial_{j}\mathbf{\boldsymbol{\Sigma}}\right)\label{eq:MAG non-conservative field}
\end{equation}
(upon removing some terms that do not contribute to their circulations)
and require the first and second spatial derivatives of the vector
$\boldsymbol{\Sigma}(\mathbf{x})$, along with the momentum fields
$\pi_{j}$ and $w_{j}$. These latter fields are the real and imaginary
parts, respectively, of the complex-valued momentum
\begin{equation}
\Pi_{j}(\mathbf{x})=\frac{\boldsymbol{\Psi}^{\dagger}(\mathbf{x})\hat{p}_{j}\boldsymbol{\Psi}(\mathbf{x})}{\boldsymbol{\Psi}^{\dagger}(\mathbf{x})\boldsymbol{\Psi}(\mathbf{x})}\label{eq:momentum field}
\end{equation}
which, written in this way, requires just the canonical momentum,
with no reference to any phase choice for the local electronic wavefunction
{[}Note that $\boldsymbol{\Psi}(\mathbf{x})$ and $\hat{p}_{j}$ are
\emph{gauge} invariant for the full \emph{e-n} problem, and that $\hat{p}_{j}$
gives the mechanical momentum when averaged over the electronic state,
$\hat{\pi}_{j}=\braket{\hat{p}_{j}}_{\text{el}}${]}. 

Note that Eq. \ref{eq:momentum field} could be also used to compute
directly the geometric phase with the help of Eq. \ref{eq:definition of QHD phase},
with results equivalent to those obtained from Eq. \ref{eq:s-expression for Berry-phase}.
However, this approach does not give full access to the electronic
properties, specifically the electromotive forces and the fields of
Eqs. \ref{eq:NBO non-conservative field}, \ref{eq:EL non-conservative field}
and \ref{eq:MAG non-conservative field} generating them. Furthermore,
the polarization vector is also useful to compute the local adiabatic
populations $p_{\pm}(\mathbf{x})=(1+\mathbf{s}_{\pm}\mathbf{s})/2$
(where $\mathbf{s}_{\pm}$ are the polarization for the $\pm$ adiabatic
states, i.e. $\mathbf{s}_{\pm}=\pm\hat{\mathbf{r}}$), from which
the adiabatic populations follow as 
\[
P_{\pm}=\int d\mathbf{x}n(\mathbf{x})\frac{1+\mathbf{s}_{\pm}\mathbf{s}}{2}
\]

\subsection{Numerical applications}

As mentioned above, in the numerical application we focused on a 2-state
2-dimensional problem by selecting the scalar $A(\mathbf{x})$ to
be a function of $x,y$ only. We set 
\[
A(\mathbf{x})=\frac{1}{2}M\omega_{x}^{2}x^{2}+\frac{1}{2}M\omega_{y}^{2}y^{2}
\]
with parameters typical of a molecular problem (the system mass $M=1$
u.m.a. and $\omega_{x}=\omega_{y}=\omega=4.5563359\times10^{-3}$
a.u., corresponding to $1000$ cm$^{-1}$), while for the effective
magnetic field we used a linear-vibronic coupling form 
\[
\mathbf{B}=\kappa_{x}x\mathbf{e}_{1}+\text{\ensuremath{\kappa}}_{y}y\mathbf{e}_{2}
\]
with $\kappa_{x}=\kappa_{y}=\kappa=0.1$ a.u.. The problem represents
thus a linear $E\otimes e$ Jahn-Teller model, with adiabatic states
$E_{\pm}(\mathbf{x})=\frac{1}{2}M\omega^{2}\rho^{2}\pm\kappa\rho$. 

We solved the time-dependent Schr\"{o}dinger equation (TDSE) using
a standard Split-Operator (SO) algorithm in conjunction with Fast-Fourier-Transforms
(FFTs) to go back and forth between real- and momentum-space. The
wavefunction was represented on a fine grid (1024$\times$1024) centered
at the CI point at $\mathbf{x}=\mathbf{0}$ and of length $20\,a_{0}$
along each direction. A small time step of $\Delta t=0.1\,$a.u. was
adopted to ensure a good sampling of the geometric phase over time
and an accurate solution of the TDSE. We considered two initial wavefunctions,
obtained by combining a ``nuclear'' wavepacket ($\psi_{0}(\mathbf{x})$)
with two (slightly) different electronic states ($\chi_{\sigma}(\mathbf{x})$),
i.e. $\Psi_{\sigma}(\mathbf{x})=\psi_{0}(\mathbf{x})\chi_{\sigma}(\mathbf{x})$.
The nuclear wavepacket $\psi_{0}(\mathbf{x})$ was a Gaussian centered
at $x_{0}=-2\kappa/M\omega^{2}$ and $y_{0}=0$, with ground-state
width along both $x$ and $y$ ($\Delta x=\Delta y=\sqrt{\hbar/2m\omega})$
and zero nominal momentum along both directions. As for the electronic
state, on the other hand, we considered both the true adiabatic ground-state
of Eq. \ref{eq:true adiabatic state} with the \emph{gauge} \emph{$\boldsymbol{\chi}=[-e^{-i\phi},1]^{t}/\sqrt{2}$
}(where $\phi$ is, as usual, the azimuthal angle of the position
vector $\mathbf{x}$) and the uniform state defined by the electronic
ground state at the center of the initial nuclear wavepacket. In other
words, we choose $\ket{u(\mathbf{x})}=\frac{1}{\sqrt{2}}(-e^{-i\phi}\ket{1}+\ket{2})$
in the first, ``correlated'' case and $\ket{u(\mathbf{x})}\equiv\ket{u_{-}(x_{0},y_{0})}$
in the second, ``uncorrelated'' case. The latter is a popular (pragmatical)
representation of the electronic ground-state, but it does not correspond
to a true adiabatic state. The difference between the two is globally
minimal (if the wavepacket is narrow enough) but evident: the first
is current-carrying even if the nuclear wavepacket is chosen real
as described above (it presents a non-vanishing velocity directed
along the negative $\phi$ direction, although very small with the
chosen parameters), the second presents a non-vanishing population
on the excited state and has a vanishing velocity field. Importantly,
the first displays a non-trivial, topological Berry phase ($+\pi$)
for any loop encircling once the origin, inherited from the adiabatic
connection, while the second has a trivial connection.

\begin{figure}[t]
\begin{centering}
\includegraphics[clip,width=1\columnwidth]{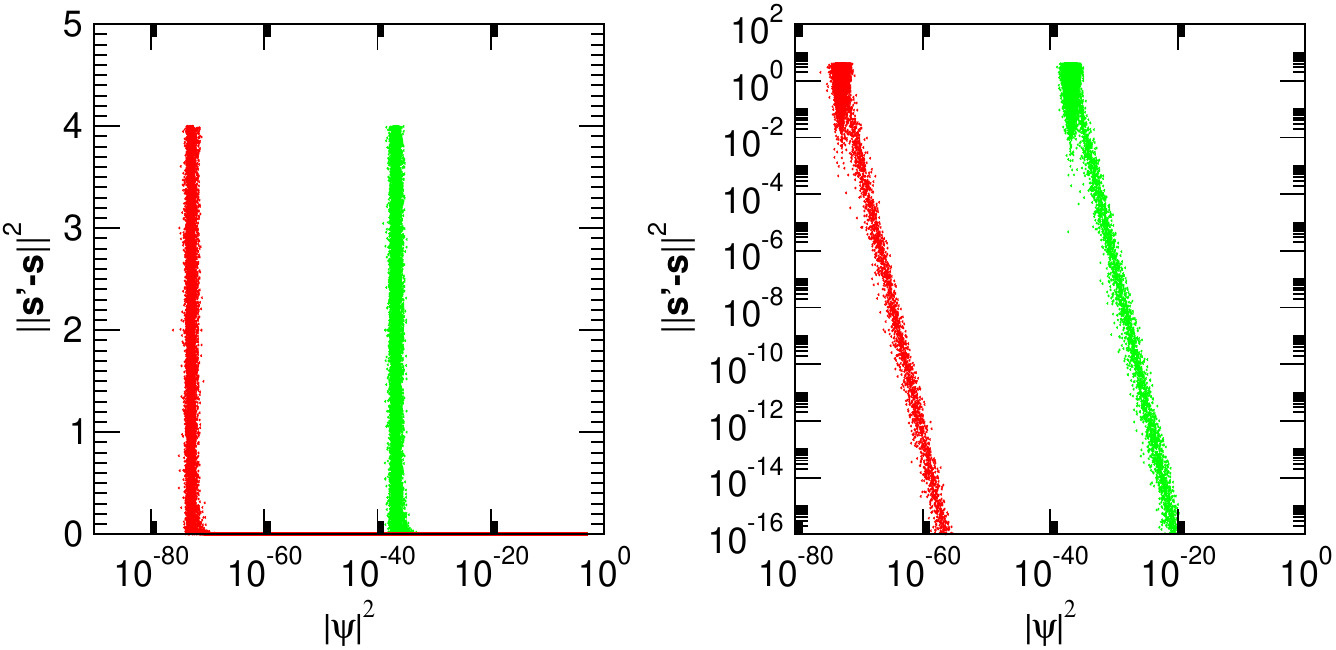}
\par\end{centering}
\caption{\label{fig:Intrinsic FFT error}Squared error intrinsic in FFT arithmetics
when computing the polarization field $\mathbf{s}(\mathbf{x})$, as
a function of the nuclear density at the grid points (see text for
details). Green and red symbols for double and quadruple precision
arithmetics, respectively.}
\end{figure}
We computed the geometric phase along selected paths at each time-step,
using the discretized version of Eq. \ref{eq:s-expression for Berry-phase},
namely
\[
\text{\ensuremath{\Gamma}}_{O}=-\frac{1}{2}\sum_{i=1}^{N_{\gamma}}\sqrt{\frac{1-s_{z}(\mathbf{x}_{i})}{1+s_{z}(\mathbf{x}_{i})}}\frac{(\mathbf{e}_{3}\times\mathbf{s}(\mathbf{x}_{i}))}{\sqrt{s_{x}^{2}(\mathbf{x}_{i})+s_{y}^{2}}(\mathbf{x}_{i})}\Delta\mathbf{s}_{i}
\]
where the $\mathbf{x}_{i}$'s are the $N_{\gamma}$ points on the
real-space grid that are used to represent the desired path $\gamma$,
$\mathbf{s}(\mathbf{x})$ is the instantaneous polarization field
and $\Delta\mathbf{s}_{i}=(\mathbf{s}(\mathbf{x}_{i+1})-\mathbf{s}(\mathbf{x}_{i-1}))/2$
(with $\mathbf{x}_{N_{\gamma}+1}=\mathbf{x}_{1}$ and $\mathbf{x}_{N_{\gamma}}=\mathbf{x}_{0}$
for a closed path). The chosen paths are circles centered at the origin
with different radii, encompassing the inner and the outer classical
turning points of the ground-state valley. 

\begin{figure}
\begin{centering}
\includegraphics[clip,width=0.9\columnwidth]{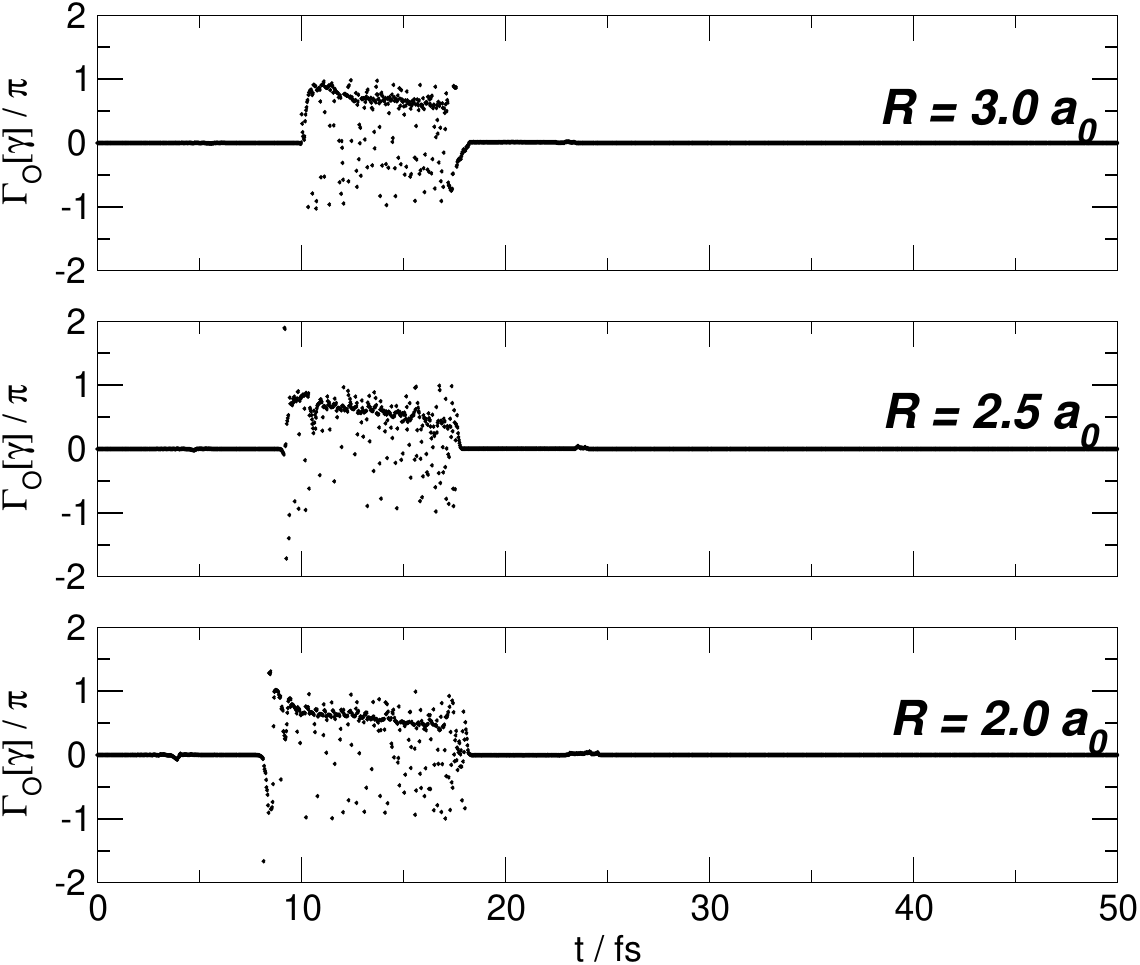}
\par\end{centering}
\caption{\label{fig:Error-estimate-in-Berry-phase}Estimated error in the computed
geometric phase. Difference between results obtained in double precision
arithmetics and $\epsilon_{\text{th}}=10^{-20}$, and results obtained
in quadruple precision arithmetics and $\epsilon_{\text{th}}=10^{-30}$.}
\end{figure}

Calculation of polarization field $\mathbf{s}(\mathbf{x})$ was critical
at the beginning of the dynamics, when the nuclear density is small
on a large portion of the grid and conflicts with the accuracy of
the numerical implementation, in particular the arithmetics underlying
the FFTs. This can be easily checked with a fake propagation step
using $\Delta t=0$ \textendash{} which only involves a forward and
a backward Fourier transformation \textendash{} by comparing the polarization
field $\mathbf{s}'(\mathbf{x})$ computed after the transformation
with that prior to the transformation ($\mathbf{s}(\mathbf{x})$).
The results of such test are shown in Fig. \ref{fig:Intrinsic FFT error},
which reports the squared error $||\mathbf{s}'-\mathbf{s}||^{2}$
at $\mathbf{x}$ as a function of the nuclear density $n(\mathbf{x})$,
for the true adiabatic state defined by Eq. \ref{eq:true adiabatic state}.
This problem has no physical implications for the geometric phase
since this phase has no physical meaning when the probe path lies
entirely or partially in a region of small nuclear density (i.e.,
where the system has little probability to be found). Nevertheless,
in order to alleviate it, we decided to update the polarization field
only at those grid points where the nuclear density was larger than
a numerically reasonable threshold $\epsilon_{\text{th}}$, and used
quadruple precision to set this threshold to $10^{-30}$ (Fig. \ref{fig:Intrinsic FFT error}).
To this end we relied on the Fortran 2003 interface to the FFTW support
for the nonstandard\texttt{ \_float128} quadruple-precision type provided
by \texttt{gcc} (http://www.fftw.org/). The results show a marginal
difference only from those obtained using double precision arithmetics
and a threshold of $10^{-20}$: as seen in Fig. \ref{fig:Error-estimate-in-Berry-phase}
the main differences arise for $t\lesssim20\,fs$, that is, when the
wavepacket has yet a tiny weight on the initially unoccupied portion
of the valley (which is accounted differently with the two thresholds
mentioned above). At longer times, precision does not affect anymore
the phase value. 
\begin{figure}
\begin{centering}
\includegraphics[clip,width=0.9\columnwidth]{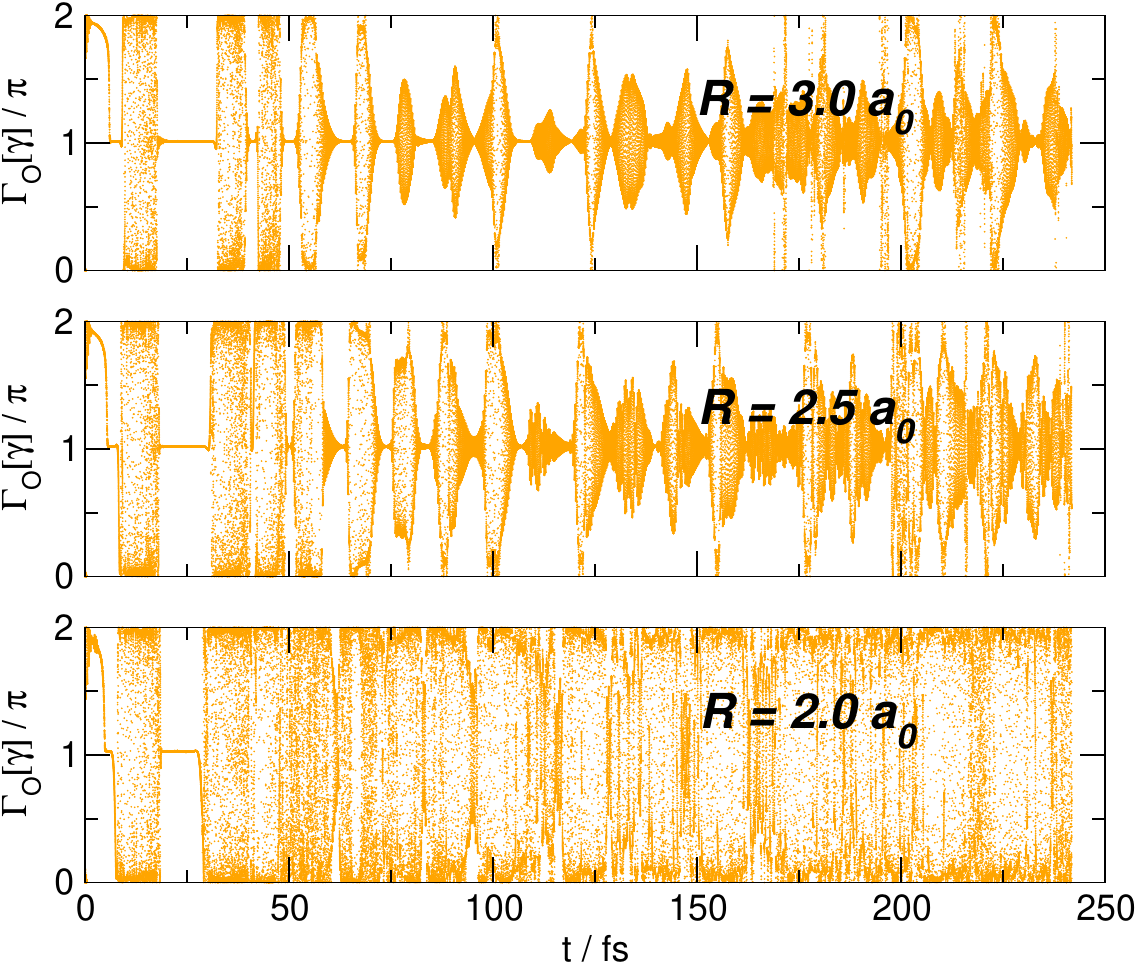}
\par\end{centering}
\caption{\label{fig:Uncorrelated initial state}Same as in Fig. \ref{fig:Exact-quantum-dynamical results}(e)
but for an uncorrelated initial state approximating the true ground
adiabatic state. }

\end{figure}
\begin{figure}
\begin{centering}
\includegraphics[width=0.9\columnwidth]{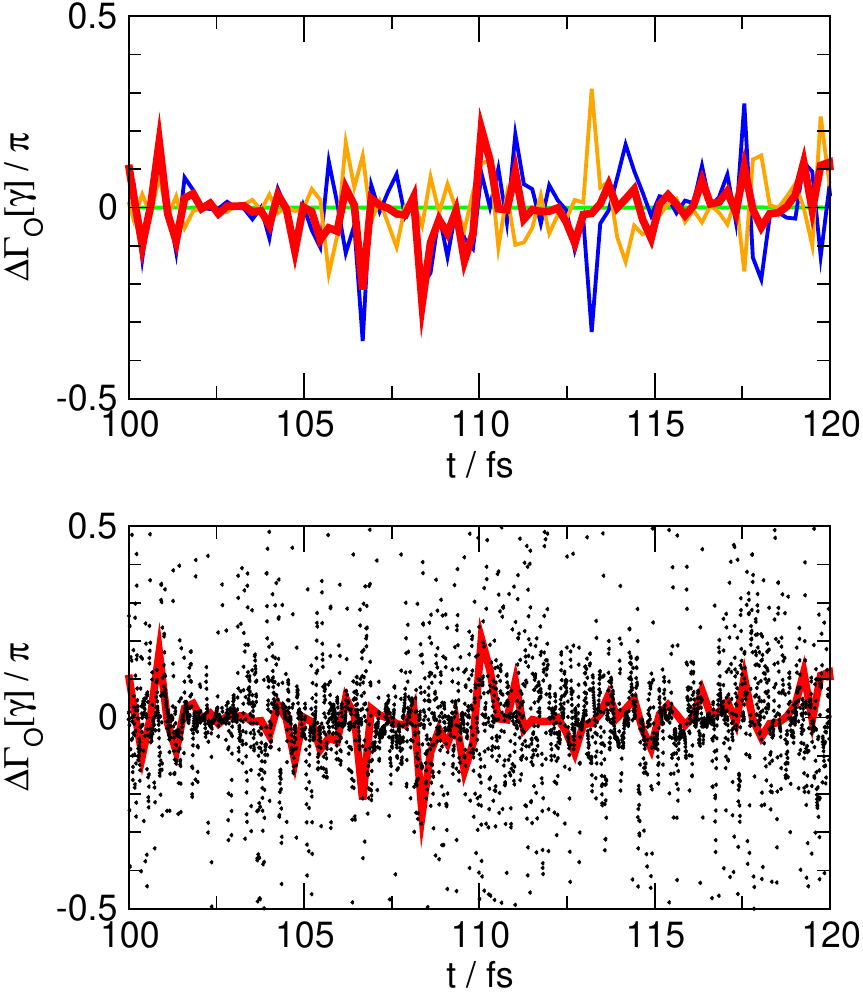}
\par\end{centering}
\caption{\label{fig:Electromitive force}Decomposition of the dynamically-induced
geometric phase change in terms of its contributing components. $\Delta\Gamma_{O}[\gamma]$
is the phase difference accumulated in a time step of $\Delta t=1$
a.u., for a circle of radius $R=1\,a_{0}$ centered at the CI point,
and an initial correlated ground-state. Bottom panel: the (minus)
total electromotive force (red) is compared with the difference $\Gamma_{O}[\gamma](t)-\Gamma_{O}[\gamma](t-\Delta t)$
in the phase computed at neighboring times (dots). Top panel: the
same total electromotive force of the bottom panel is reported along
with the NBO (green), \emph{pseudo}-electric (blue) and \emph{pseudo}-magnetic
(orange) components. }

\end{figure}

The evolution of the geometric phase for the correlated initial state
are discussed in the main text, see Fig. \ref{fig:Exact-quantum-dynamical results}.
The uncorrelated initial state gives rise to qualitatively similar
results but presents a more marked non-adiabatic behaviour (see Fig.
\ref{fig:Uncorrelated initial state}), in accordance with an excited
probability which is two order of magnitude larger than in the previous
case, $P_{+}\approx10^{-3}$. Noteworthy, despite the evident variations
the phase oscillates around the adiabatic value $\pi$ and gets back
to this value at regular time intervals. 

Finally, we further computed the total electromotive force and their
contributing non-Born-Oppenheimer, \emph{pseudo}-electric and \emph{pseudo}-magnetic
components. The results are shown in Fig. \ref{fig:Electromitive force}
for a circle of small radius, $R=1.0\,a_{0}$, where these forces
are more evident, for the correlated initial state. The electromotive
force and its contributing components were obtained from Eqs. \ref{eq:NBO non-conservative field},
\ref{eq:EL non-conservative field} and \ref{eq:MAG non-conservative field},
at intervals much larger than the time step used for the dynamics
in order to reduce the computational cost (Eqs. \ref{eq:NBO non-conservative field}-\ref{eq:MAG non-conservative field}
require the first and second spatial derivatives of $\mathbf{s}$,
hence add FFT calls to the propagation). As seen from that figure,
the NBO component contributes little to the phase change, while the
\emph{pseudo}-electric one plays the major role. This implies that
these dynamically-induced phase changes cannot be captured by approximate
methods like Ehrenfest dynamics which only account for the NBO term.

\end{document}